\documentclass{iopart}
  \expandafter\let\csname equation*\endcsname\relax
  \expandafter\let\csname endequation*\endcsname\relax
\usepackage{graphicx}
\usepackage{amsmath}
\usepackage{cite}
\usepackage{amssymb}
\usepackage{caption}
\usepackage{subcaption}

\DeclareMathOperator{\sinc}{sinc}
\DeclareMathOperator{\erf}{Erf}

\newcommand{\expect}[1]{\ensuremath{\left<#1\right>}}
\newcommand{\ket}[1]{\ensuremath{\left|#1\right>}}

\renewcommand{\Im}{\operatorname{Im}}

\begin{document}

\title{Squeezing in Bose-Einstein condensates with large numbers of atoms}

\author{Mattias T Johnsson$^1$, Graham R Dennis$^2$ and Joseph J Hope$^1$}

\address{$^1$Department of Quantum Science, Research School of Physics and Engineering, The Australian National University, Canberra ACT 0200, Australia}
\address{$^2$Plasma Research Laboratory, Research School of Physics and Engineering, The Australian National University, Canberra ACT 0200, Australia}
\ead{mattias.johnsson@anu.edu.au}

\begin{abstract}
We examine the feasibility of creating and measuring large relative number squeezing in multicomponent trapped Bose-Einstein condensates.  In the absence of multimode effects, this squeezing can be arbitrarily large for arbitrarily large condensates, but a range of processes limit the measurable squeezing in realistic trap configurations.  We examine these processes, and suggest methods to mitigate them. We conclude that high levels of squeezing with large numbers of atoms is feasible, but can realistically only be achieved in particular trap geometries.  We also introduce a method of maximising the measurable squeezing by using a $\pi$-pulse during the process to improve spatial mode-matching.
\end{abstract}

\maketitle

\section{Introduction}
\label{sectionIntroduction}

Quantum degenerate ultracold atomic sources have been used to study fundamental processes such as quantum phase transitions \cite{Leggett2001,Buluta2009}, quantum non-equilibrium thermodynamics \cite{Kinoshita:2006,Barnett:2011} and entanglement of massive particles \cite{Kheruntsyan:2005}.  Their ability to be highly isolated from their environment and controlled precisely by optical, magnetic and rf fields makes them an ideal platform for engineering and manipulating non-trivial quantum states. In quantum optics, the most common way to produce entangled states has been through squeezing, and it has been recognised that the intrinsic nonlinearities of atoms naturally lead to quadrature squeezing \cite{Kheruntsyan:2005, johnssonET2007, haineET2009, doeringET2010}.  Number squeezing can be achieved with spin squeezing techniques \cite{kitagawaET1993, gross2010, sinatra2012}, molecular dissociation of Bose-Einstein Condensates (BECs) \cite{kheruntsyanET2002,davisET2008,savage2007,ogren2010}, four-wave mixing \cite{perrin2008, ogren2009, jascula2010}, double-well potentials \cite{esteve2008, maussang2010}, or mixing quadrature-squeezed states using Ramsey interferometry \cite{haineET2009}. The creation of these non-classical atomic states has relevance for fundamental tests of quantum mechanics such as the EPR paradox \cite{ogren2010, pu2000, sorensen2001, haine2005,zhao2007}, as well as atom interferometry \cite{doeringET2010, gross2010, sorensenET1999, kuzmichET2000, meyerET2001, liebfriedET2004, roosET2006}.
While these techniques have resulted in significant experimentally detectable number squeezing, they have been limited to small numbers of atoms, typically a few hundred \cite{gross2010, esteve2008}. 

Ultracold atomic interferometers are also reaching state of the art sensitivities in precision measurements \cite{robins:2013, altinET2013}. While all current atom interferometry is limited by technical noise sources, it appears that BEC-based coherent sources offer the best chance of overcoming some of these issues, for the same reasons that optical lasers are typically preferred for optical interferometry. In particular, it appears that Bose-condensed sources allow for higher fidelity mirrors and beamsplitters \cite{szigeti:2012}, are less sensitive to the effects of distortions in the optical wavefronts of the beamsplitters and mirrors and the Coriolis effect \cite{Debs:2011, robins:2013}, and are robust to the adverse effects of strong outcoupling when feedback-cooled \cite{semiclassicalstabilization, Szigeti:2009, Szigeti:2010}. 

If these technical noise sources can be reduced and/or controlled, atom interferometers will be limited by atomic shot noise, which leads to a sensitivity that scales as the square root of the atomic flux. From that point, sensitivity will only be improved by increasing the atomic flux and brightness \cite{semiclassicallimit2007, outcouplercomparison, jeppesenET2008}, and/or using non-classical quantum states such as squeezed states to go below the shot noise limit \cite{robins:2013}. 

Producing squeezed states for metrological purposes is therefore only relevant when it can be achieved in a context of large numbers of atoms.  This paper examines the generation of atomic squeezing using the inherent Kerr-like nonlinearities of ultracold atoms, and demonstrates a trap geometry in which metrologically relevant levels of squeezing are present in trap geometries that are compatible with large atomic number.

The creation, detection and application of squeezing requires precise mode-matching, and is therefore very sensitive to uncontrolled coupling between spatial modes of the atomic field.  Atomic squeezing experiments have typically attempted to operate in the single-mode regime, where the trap is tight enough such that all higher modes are frozen out \cite{LiET2009}.  However, tight traps limit the total number of atoms, as the 3-body recombination rate places an effective maximum atomic density in the gas, and well before that limit, the interatomic interactions break the single-mode operation.  Both of these effects limit the number of atoms that can be squeezed effectively, so we will examine a broader range of trap geometries, and calculate the dynamics in the presence of multiple spatial modes. Many similar previous simulations have been performed in one or two spatial dimensions, by integrating out some of the remaining dimensions \cite{Kheruntsyan:2005, johnssonET2007, haineET2009, kheruntsyanET2002}.  For simulations of semiclassical fields this can be a reasonable approximation, but it can ignore spontaneous scattering processes in full quantum field theory calculations, so we take care to include full 3D simulations. 

In Section \ref{secTwoModeAnalytic} we examine the ideal behaviour of a nonlinear bosonic two-state system when that system possesses only a single spatial mode, extracting full analytic solutions for the number squeezing in each of the two states, as well as the number difference squeezing between the two states. These solutions of the ideal case provide a check on the multimode cases we consider later in the paper, as well as providing a bound on the best possible squeezing for the system.  In Section \ref{sec:MMdescription} we show how multimode behaviour becomes important in physical systems, and introduce our model for simulating higher dimensional trap geometries.  Section~\ref{sec:1Dsim} analyses one-dimensional models, highlighting some of the issues regarding optimising squeezing parameters and mode-matching in squeezing experiments.  Section~\ref{sec:pipulse} shows that the application of a $\pi$-pulse during the process can significantly improve mode-matching issues.  Section~\ref{sec:2D3Dsqueezing} investigates measurable squeezing in 2D and 3D, and uses a Bogoliubov analysis to explain the scaling with system size and dimensionality. Section~\ref{sectionConclusion} describes the conclusion that large volume, flat-bottomed, three-dimensional traps, which are obviously compatible with large atom number, can produce extremely high degrees of squeezing.

\section{Model and single-mode solutions}
\label{secTwoModeAnalytic}
Our goal is to describe the number squeezing possibilities of a BEC comprising two relevant internal states denoted by $|a\rangle$ and $|b\rangle$.  The Hamiltonian for a pair of coupled atomic internal modes is the spatial integral of the Hamiltonian density:
\begin{equation}
\hat{\mathcal{H}} = \sum_{j\in \{a,b\}} \hat{\psi}_j^{\dagger}\left(-\frac{\hbar^2}{2 m}\nabla^2+V_j\right)\hat{\psi}_j 
          + \sum_{i,j}\frac{U_{i j}}{2} \hat{\psi}_i^{\dagger} \hat{\psi}_j^{\dagger} \hat{\psi}_j \hat{\psi}_i
          + \kappa \hat{\psi}_a^{\dagger} \hat{\psi}_b + \kappa^* \hat{\psi}_b^{\dagger}  \hat{\psi}_a,
\label{eqFieldHamiltonian}
\end{equation}
where $\hat{\psi}_j(\mathbf{r})$ is the annihilation field operator for a particle at position $\mathbf{r}$ and in internal state $|j\rangle$, $m$ is the mass of the atoms, $V_j(\mathbf{r})$ is the trapping potential for atoms in internal state $|j\rangle$, $\kappa$ describes the coupling between the two states, and the $U_{ij}$ describe the various inter- and intra-state nonlinearities and are given by
\begin{equation}
U_{ij} = \frac{4 \pi \hbar^2 a_{ij}} {m},
\label{eqUij}
\end{equation}
where $a_{ij}$ are the inter- and intra-species s-wave scattering lengths.

We begin by considering a simplified version of the problem, assuming that only one of these spatial modes is relevant for each of the condensate's components.  This is done by writing the multimode atomic field operators in terms of a set of basis functions $\hat{\psi}_a({\mathbf{r}},t) = \sum_n u_{n}({\mathbf{r}}) \hat{a}_{n}(t)$ and $\hat{\psi}_b({\mathbf{r}},t) = \sum_n u_{n}({\mathbf{r}}) \hat{b}_{n}(t)$, where $\hat{a}_{n}$ and $\hat{b}_{n}$ annihilate a particle with normalised spatial mode $u_n({\mathbf{r}})$ and internal state $|a\rangle$ or $|b\rangle$ respectively. With this expansion, the nonlinear term in the multimode Hamiltonian for the self-interaction of the $|a\rangle$ internal state becomes
\begin{equation}
\hat{H} = \frac{U_{aa}}{2} \int dV \sum_{nmpq} u_n^* u_m^* u_p u_q \, \hat{a}^{\dagger}_{n} \hat{a}^{\dagger}_{m} \hat{a}_{p} \hat{a}_{q} .
\end{equation}
When the system remains predominantly in spatial mode $u_0(\mathbf{r})$, we can approximate this sum as the single term $\hat{H} = \hbar\chi_{aa} \hat{a}^{\dagger} \hat{a}^{\dagger} \hat{a} \hat{a}/2$ where
\begin{equation}
\chi_{aa} = \frac{1}{\hbar} U_{aa} \int |u_{0}({\mathbf{r}})|^4 \, dV.
\label{eqChiUequivalence}
\end{equation}
We proceed similarly for the other terms.  In the parameter regimes we are interested in, the detuning between the two spatial modes will be irrelevant, so setting our zero of energy appropriately, we are left with only a standard Kerr-type nonlinearity arising from atom-atom interactions, as well as a linear coupling between the two modes.  The Hamiltonian governing this reduced system is given by
\begin{equation}
\hat{H} = \hbar\frac{\chi_{aa}}{2} \hat{a}^{\dagger} \hat{a}^{\dagger} \hat{a} \hat{a}
          + \hbar\chi_{ab} \hat{a}^{\dagger} \hat{a} \hat{b}^{\dagger} \hat{b}
          + \hbar\frac{\chi_{bb}}{2} \hat{b}^{\dagger} \hat{b}^{\dagger} \hat{b} \hat{b}
          + \hbar\Omega (\hat{a}^{\dagger} \hat{b} + \hat{b}^{\dagger}  \hat{a} ),
\label{eqTwoModeHamiltonian}
\end{equation}
where the $\chi_{ij}$ describe the various inter- and intra-mode nonlinearities, and $\Omega$ couples the two modes and allows for population transfer between them.

When considering squeezing, the Hamiltonian (\ref{eqTwoModeHamiltonian}) has typically been handled within the framework of spin squeezing, which has a long history \cite{kitagawaET1993, esteve2008,sorensenET1999, liebfriedET2004, wineland1994, steel+collett}. This approach involves exploiting the equivalence between the algebra of two harmonic oscillators and that of angular momentum, and writing
\begin{eqnarray}
\hat{J}_+ &=& \hat{a}^{\dagger} \hat{b}, \\
\hat{J}_- &=& \hat{a} \hat{b}^{\dagger}, \\
\hat{J}_x &=& \frac{1}{2} \left( \hat{J}_+ + \hat{J}_-\right), \\
\hat{J}_y &=& \frac{1}{2i} \left( \hat{J}_+ - \hat{J}_-\right), \\
\hat{J}_z &=& \frac{1}{2} \left( \hat{a}^{\dagger} \hat{a} - \hat{b}^{\dagger} \hat{b} \right).
\end{eqnarray}
Using these variables Eq.~(\ref{eqTwoModeHamiltonian}) becomes
\begin{equation}
\hat{H} = \hbar\Delta \omega(\hat{N}) \, \hat{J}_z + \frac{1}{2}\hbar\chi_+ \hat{J}_z^2 + 2\hbar\Omega \hat{J}_x - \frac{1}{4}\left(\chi_{aa} + \chi_{bb}\right)\hat{N} + \frac{1}{8} \left(\chi_{aa} + \chi_{bb} + 2\chi_{ab}\right) \hat{N}^2,
\label{eqNonlinearJosephson}
\end{equation}
where $\hat{N}=\hat{a}^\dagger\hat{a} + \hat{b}^\dagger\hat{b}$ is the total number operator, $\Delta \omega(\hat{N}) = (\hat{N}-1)\left(\chi_{aa} - \chi_{bb}\right)/2$ and $\chi_+=\chi_{aa}+\chi_{bb}-2\chi_{ab}$. For initial states which are total number Fock states, the last two terms are physically meaningless phase shifts and the first term can be removed by an appropriate choice of detuning.  In this case Eq.~\eqref{eqNonlinearJosephson} reduces to the Josephson Hamiltonian with zero detuning \cite{steel+collett}
\begin{equation}
  \hat{H}_J = \frac{1}{2} \hbar\chi_+ \hat{J}_z^2 + 2\hbar \Omega \hat{J}_x. \label{eqJosephson}
\end{equation}
The system described by $\hat{H}_J$ is then solved using angular momentum algebra.

The reduction of Eq.~\eqref{eqTwoModeHamiltonian} to the Josephson Hamiltonian has led to the common misconception that no squeezing can be generated in this system if $\chi_+ = 0$ as the effects of the $\Delta\omega$ term are usually dismissed as an irrelevant detuning.  This is not true in general for initial states other than a total number Fock state as the $\Delta\omega(\hat{N})\hat{J}_z$ term leads to a number-dependent detuning.  We demonstrate in this paper that for the case of an initial coherent state this term leads to squeezing.  A spin-echo pulse applied half-way through the experiment which exchanges the population of modes $\ket{a}$ and $\ket{b}$ (which is the protocol used in most spin squeezing experiments) also cancels the effects of this term, and in this case squeezing is only generated if $\chi_+ \neq 0$.  In this work we directly solve Eq.~\eqref{eqTwoModeHamiltonian} demonstrating that squeezing is generated (in the absence of a spin-echo pulse) even if $\chi_+=0$.


The system is prepared with all the population initially in $|a\rangle$, and vacuum in $|b\rangle$. We choose the initial state $|a\rangle$ to be a coherent state with average number $\langle \hat{a}^{\dagger} \hat{a} \rangle = N$.  As the system is insensitive to the initial phase of the coherent state, this is equivalent to a mixture of coherent states of uncertain phase, or a mixture of Poisson-distributed number states. Such a state is consistent with BEC coherence experiments \cite{Hadzibabic2004}.  

\begin{figure}
    \centering
    \includegraphics[width=8cm]{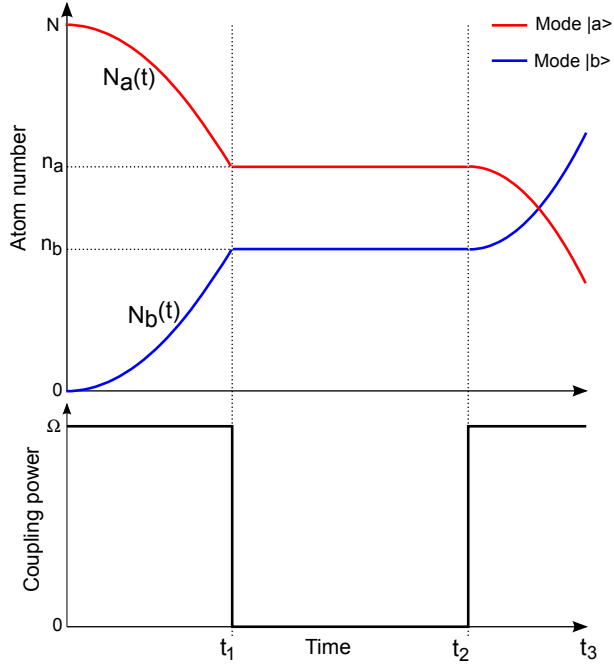}
    \caption{Schematic of the pulse sequence and atomic populations as a function of time.  Typically the coupling sequences at the beginning and end of the sequence are much shorter than the hold time, $\tau_{\mathrm{hold}} = t_2 - t_1 $, which is where the majority of the squeezing occurs.}
    \label{figPulseScheme}
\end{figure}

At time $t=0$, the coupling $\Omega$ is applied until time $t=t_1$, resulting in a portion of the population being transferred into mode $|b\rangle$. We assume this transfer process is fast relative to the nonlinear energy time scale, ensuring that, as the initial states were coherent states, after the population transfer both $|a\rangle$ and $|b\rangle$ are still described by coherent states with mean number $\langle \hat{a}^{\dagger} \hat{a} \rangle = n_a$ and $\langle \hat{b}^{\dagger} \hat{b} \rangle = n_b$ respectively.  The coupling $\Omega$ is then switched off until time $t_2$, while the atoms interact solely through the nonlinear terms. After this hold time $\tau_{\mathrm{hold}} = t_2 - t_1 $, the coupling $\Omega$ is switched back on until time $t_3$, with a phase shift $\phi$ compared to its first application. During this last stage, the two modes exchange population and the quadrature variances are converted into number variance, in the same way that homodyne measurements are used in quantum optics to convert quadrature squeezing into number squeezing, which can be directly measured. In this interpretation, $\phi$ is the relative phase of the strong local oscillator, which allows specific phase angles of the quadrature squeezing to be examined.  This experiment can also be interpreted as a Ramsey interferometer with a final beam splitter phase of $\phi$.  The entire sequence of pulses and the resulting populations are shown schematically in Figure~\ref{figPulseScheme}.

We solve this system analytically, and derive expressions for absolute and relative number squeezing. We present the solutions here; the full derivation can be found in \ref{appendixTwoModeDerivation}. If we define
\begin{eqnarray}
s &=& \sin(\theta), \label{eqsDef} \\
c &=& \cos(\theta), \\
\theta &=& \Omega(t_3 - t_2), \\
\lambda_{ij} &=& \chi_{ij} \tau_{\mathrm{hold}}, \\
A &=& \sqrt{n_a} \exp [n_a (e^{-i(\lambda_{aa}-\lambda_{ab})} - 1 )], \\
A_2 &=& n_a \exp [n_a (e^{-2i(\lambda_{aa}-\lambda_{ab})} - 1 )], \\
B &=& -i e^{i \phi}\sqrt{n_b} \exp [n_b (e^{-i(\lambda_{bb}-\lambda_{ab})} - 1 )], \\
B_2 &=& -n_b e^{2i\phi} \exp [n_b (e^{-2i(\lambda_{bb}-\lambda_{ab})} - 1 )], \\
D &=& AB^* - A^*B, \label{eqDDef}
\end{eqnarray}
then the expectation values for number, number variance, and number difference variance after applying the final coupling field are given by
\begin{eqnarray}
N_a(t_3) &=& n_a c^2 + n_b s^2 + i c s D, \\
N_b(t_3) &=& n_a s^2 + n_b c^2 - i c s D, \\
{\mathrm{Var}} [ N_a](t_3) &=& n_a c^2 + n_b s^2 + 2 n_a n_b c^2 s^2 \nonumber\\
       && + c^2 s^2 (D^2 - e^{i(\lambda_{aa} - \lambda_{bb})} B_2 A_2^* - e^{-i(\lambda_{aa} - \lambda_{bb})} B_2^* A_2) \nonumber \\
       && + i c s D (1-2 n_a c^2 -2 n_b s^2)  \nonumber\\
       && + 2 i c^3 s n_a (e^{-i(\lambda_{aa} - \lambda_{ab})} A B^* - e^{i(\lambda_{aa} - \lambda_{ab})} A^* B ) \nonumber \\
       && + 2 i c s^3 n_b (e^{i(\lambda_{bb} - \lambda_{ab})} A B^* - e^{-i(\lambda_{bb} - \lambda_{ab})} A^* B ), \\
{\mathrm{Var}} [ N_b](t_3) &=&  n_a s^2 + n_b c^2 + 2 n_a n_b c^2 s^2 \nonumber \\
       && + c^2 s^2 (D^2 - e^{i(\lambda_{aa} - \lambda_{bb})} B_2 A_2^* - e^{-i(\lambda_{aa} - \lambda_{bb})} B_2^* A_2 ) \nonumber \\
       && + i c s D (-1+2 n_a s^2 + 2 n_b c^2)  \nonumber \\
       && + 2 i c^3 s n_b (e^{-i(\lambda_{bb} - \lambda_{ab})} A^* B - e^{i(\lambda_{bb} - \lambda_{ab})} A B^* ) \nonumber \\
       && + 2 i c s^3 n_a (e^{i(\lambda_{aa} - \lambda_{ab})} A^* B - e^{-i(\lambda_{aa} - \lambda_{ab})} A B^* ), \\
{\mathrm{Var}} [ N_a - N_b](t_3) &=& n_a + n_b + 4 i c s D (n_a - n_b)(s^2 - c^2) \nonumber \\
       && + 4 c^2 s^2 (2 n_a n_b +D^2 - e^{i(\lambda_{aa} - \lambda_{bb})} A_2^* B_2 - e^{-i(\lambda_{aa} - \lambda_{bb})} A_2 B_2^* ) \nonumber \\
       && + 4 i c s (c^2 - s^2) \left( n_a e^{-i(\lambda_{aa} - \lambda_{ab})} A B^* - n_a e^{i(\lambda_{aa} - \lambda_{ab})} A^* B \right. \nonumber \\
       && \,\,\,\,\, \left. + n_b e^{-i(\lambda_{bb} - \lambda_{ab})} A^* B - n_b e^{i(\lambda_{bb} - \lambda_{ab})} A B^* \right). \label{eqNumDiffVariance}
\end{eqnarray}
In the case where all the nonlinearities are equal, such that $\chi_{aa} = \chi_{ab} = \chi_{bb}$, the number variances all reduce to those of a coherent state, i.e. ${\mathrm{Var}}[N_a]=N_a$, and no number squeezing is possible. It should also be noted that, although there are three independent nonlinearities $\chi_{ij}$ in the Hamiltonian (\ref{eqTwoModeHamiltonian}), the expressions for the number variances only depend on differences. This means that number squeezing is parameterised by only two independent quantities, for example $\chi_{aa}-\chi_{ab}$ and $\chi_{aa}-\chi_{bb}$.  



\begin{figure}
    \centering
    \includegraphics[width=10cm]{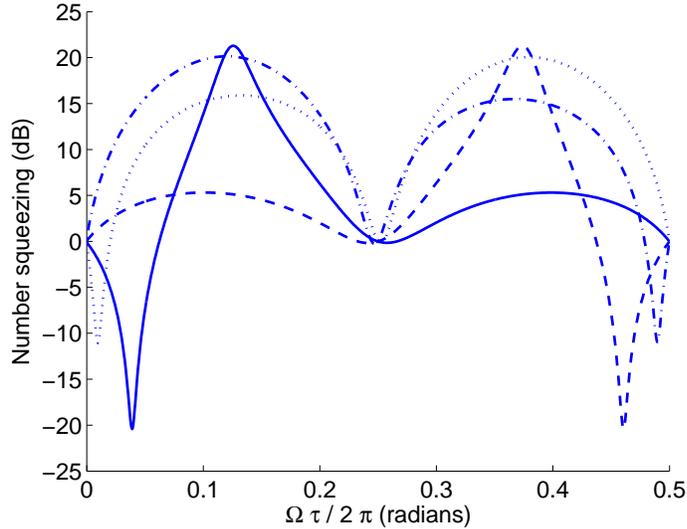}
    \caption{Normalised number variance Var$[N_a]/N_a$ in mode $|a \rangle$ as a function of final recombination time plotted on a logarithmic scale, using four different mixing angles $\phi$. Parameters: $n_a = n_b =5 \times 10^5$, $\chi_{aa}=0.04\, \text{s}^{-1}, \chi_{ab}=0, \chi_{bb}=0.01\,\text{s}^{-1}, \tau_{\mathrm{hold}}=4\times 10^{-4}$s. Solid line, $\phi=0.10$; dash-dotted line, $\phi=1.42$; dashed line, $\phi=3.24$; dotted line, $\phi=4.71$. These particular parameters yield 21\,dB of number squeezing for the $\phi=0.1$ and $\phi=1.42$ cases.} 
    \label{figTwoModeAnalyticExamples}
\end{figure}

\begin{figure}
    \centering
    \includegraphics[width=10cm]{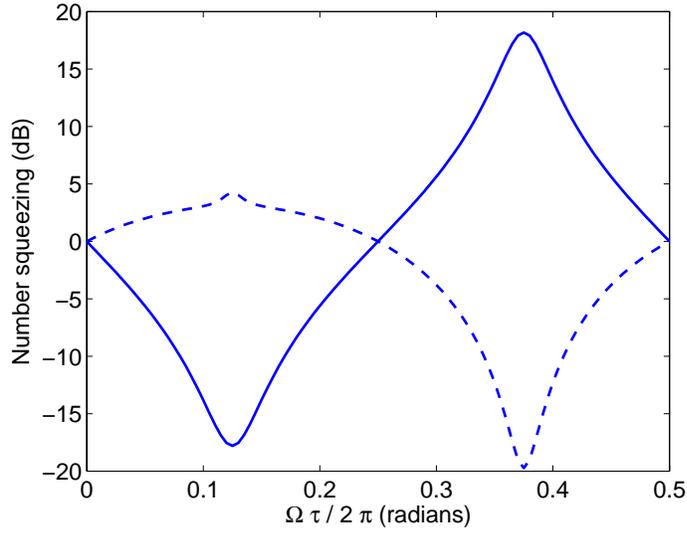}
    \caption{Number squeezing in state $|a\rangle$ (solid line) and number difference squeezing between states $|a\rangle$ and $|b\rangle$ (dashed line) as a function of final recombination time in the case where $\chi_+ = \chi_{aa} + \chi_{bb} - 2 \chi_{ab} = 0$. Parameters: $n_a = n_b = 2 \times 10^5$, $\chi_{aa}=0.03\, \text{s}^{-1}, \chi_{ab}=0.02, \chi_{bb}=0.01\,\text{s}^{-1}, \tau_{\mathrm{hold}}=2\times 10^{-3}$s. Solid line, $\phi=1.67$; dashed line, $\phi=1.55$.} 
    \label{figZeroEffectiveChi}
\end{figure}

In Figure \ref{figTwoModeAnalyticExamples} we plot a solution for the normalised number variance in mode $|a\rangle$ as a function of $\phi$, choosing a typical set of nonlinearities. This example illustrates the basic features of the system: during the final coupling period, the normalised number variance undergoes cyclic variation with a period $\tau=\pi / \Omega$, and has a series of minima. It is necessary to choose both the correct phase $\phi$ as well as the correct length of time for the coupling to be applied in order to obtain optimum squeezing.  As mentioned previously, squeezing is possible in this system even when the nonlinearity in the reduced Josephson Hamiltonian $\chi_+=\chi_{aa} + \chi_{bb} - 2\chi_{ab} = 0$.  Figure~\ref{figZeroEffectiveChi} demonstrates that significant squeezing is possible in this system if a spin-echo pulse is not used.

This two-mode analytic model allows arbitrarily good number squeezing, provided there are no limits to the number of particles in the system or to the hold time. The two-mode approximation can be valid in tight traps with very low particle number, and this result correctly predicts the squeezing demonstrated by Gross \textit{et al.} \cite{gross2010}. Unfortunately, for larger numbers of atoms in realistic traps there are more than two modes present, and these multimode effects limit the level of squeezing that can be achieved.

\section{Multimode number squeezing analysis} \label{sec:MMdescription}

The zero-dimensional, two-mode model described in the previous section shows that arbitrarily good squeezing can be achieved using the intrinsic nonlinearities in a BEC, but it ignores the existence of multiple spatial modes.  There are three problems caused by the existence of multiple modes, and these will limit the achievable squeezing.

The first problem is present even if the spatial modes are uncoupled.  The two-mode model shows that the parameters required to obtain best squeezing are dependent on the strength of the nonlinearity. In the multimode case, the effective nonlinearity of a mode is a function of the mode shape, as is shown by Eq.~(\ref{eqChiUequivalence}). The optimum hold time $\tau_{\mathrm{hold}}$ depends on the product of the effective mode nonlinearity and the number of particles in that mode, as does the recombination phase $\phi$.  Different modes in a multimode environment will typically have varying effective nonlinearity-particle number products and therefore squeeze at different rates, with each mode having a different optimum $\tau_{\mathrm{hold}}$ and optimum recombination phase $\phi$. This means that the best squeezing will be achieved by choosing values of $\tau_{\mathrm{hold}}$ and $\phi$ that, averaged across all the spatial modes present, result in the overall lowest total number variance. Clearly, however, this averaged number variance will be still be higher than the specific mode that exhibits the best squeezing.

The second problem is due to the number-dependent dynamics of the spatial modes, which is observable even in a semiclassical simulation.  Due to the nonlinear term in the Hamiltonian, the local phase evolution of the atomic field is density dependent.  Consequently the mode shape of the atomic field changes depending on the nonlinearity-density product, which means that unless $U_{aa}|\psi_a({\mathbf{r}})|^2 + U_{ab}|\psi_b({\mathbf{r}})|^2 = U_{bb}|\psi_b({\mathbf{r}})|^2 + U_{ab}|\psi_a({\mathbf{r}})|^2$ for all ${\mathbf{r}}$, the mode shapes will not overlap perfectly during the final coupling pulse. This degraded mode-matching will result in lower efficiency when converting the quadrature squeezing to number squeezing. 

We might speculate that the effects of these first two problems would be lessened by trap geometries that lead to near-constant density profiles for the BEC.  This should enable higher squeezing.  We cannot make strong conclusions without also considering the effects of coupling between spatial modes, however, which leads to the third potential problem: coupling between modes is inevitable in the presence of a nonlinearity, meaning that the modes cannot be analysed independently. Coupling between modes will disturb mode-matching, as well as mixing fields with different phase evolution, which can rapidly destroy squeezing.  

An analytic solution to the multimode, higher-dimensional problem is intractable, so we must turn to numerical solutions.  We use stochastic methods based on phase space representations to simulate the dynamics of the multimode quantum fields.  These methods alleviate the problem of directly simulating elements of the Hilbert space, the size of which increases exponentially with the number of spatial modes \cite{gardiner1991, steelET1998}.  Stochastic methods achieve this by finding a sufficiently well-behaved quasi-probability representation for the density matrix, which can then be simulated as the average behaviour of a number of low-dimensional samples.  These sample objects are the same size as the equivalent classical field.  In our case, the only tractable phase space representation is the functional Wigner representation \cite{steelET1998}, which is well behaved when the quantum field is approximately Gaussian, as it is for coherent and squeezed states.  Stochastic methods based on the functional Wigner representation have been used to model the behaviour of BECs and atom lasers \cite{johnssonET2007,dallET2009,dennisET2010}, as the Wigner representation allows the mapping of the system to a Fokker-Planck equation with positive semi-definite diffusion matrix (unlike the $P$-representation \cite{gardiner1991}), and does not suffer exponential path-weighting problems that would constrain it to very short simulation intervals (as would the positive-$P$ representation \cite{gardiner1991}).

The equation of motion for the density matrix is defined by the Hamiltonian in Eq.~(\ref{eqFieldHamiltonian}).  This defines the evolution of the functional Wigner distribution of the system, which has a one-to-one correspondence with the density matrix.  The equation of motion for the functional Wigner distribution contains derivatives of third order which we assume to be negligible.  This uncontrolled approximation is called the Truncated Wigner Approximation (TWA), and while it has been used widely on ultracold gases \cite{johnssonET2007,haineET2009,steelET1998,dennisET2010,johnssonHope:2007}, care must be taken that the simulation remains valid. For our simulations we checked the results of our simulations against the analytic two-mode solution, as well as checking for typical indications of TWA breakdown such as the appearance of negative densities in lightly populated modes.  We were also able to use the Bogoliubov analysis described in Section \ref{sec:2D3Dsqueezing} as an independent estimate of the validity of the TWA.  We found that the TWA remained valid except in some 3D simulations with high nonlinearities, where there were a large number of modes, and the number of particles per mode could drop to ten or less in the densest regions where the squeezing was generated. Over very long time scales, this low mode occupation began to result in TWA breakdown.  Fortunately, the best squeezing was typically found in regions where the TWA was valid.

Under the TWA, the equation of motion for the functional Wigner distribution is a Fokker-Planck equation, and can therefore be sampled by a set of stochastic equations, as described in \cite{gardiner1991}.  These stochastic fields $\phi(\mathbf{x})$ are related to field operator expectation values by  
\begin{equation}
\langle :\hat{\psi}^\dagger(\mathbf{x_1})\cdots\hat{\psi}^\dagger(\mathbf{x_n})\hat{\psi}(\mathbf{y_1})\cdots\hat{\psi}(\mathbf{y_m}):_{\mbox{sym}}\rangle = \mathbb{E}\left[\phi^*(\mathbf{x_1})\cdots\phi^*(\mathbf{x_n})\phi(\mathbf{y_1})\cdots\phi(\mathbf{y_m})\right],
\label{eqExpectationRelations}
\end{equation}
where $:\star:_{\mbox{sym}}$ denotes symmetric ordering of the operators, and $\mathbb{E}$ denotes a stochastic average over the variables $\phi(\mathbf{x})$.  While we see below that the evolution of these fields is deterministic, the initial state still requires a random element, so multiple realisations are required.  When approximating these equations on a discrete grid, the magnitude of initial noise, the relationship between stochastic averages and the expectation values, and the equations of motion all become grid-dependent.  This is not a signature of a fundamental problem, as the Hamiltonian in Eq.~(\ref{eqFieldHamiltonian}) assumes a contact potential between the atoms, which is not correct below a length scale that can probe the details of the true potentials.  Furthermore, it is not a problem in practice, as the calculation gives grid-independent predictions for physical observables well before that regime is reached.

These stochastic equations take the form
\begin{eqnarray}
\frac{d \phi_a}{dt} = -\frac{i}{\hbar}\left(-\frac{\hbar^2}{2 m}\nabla^2+V_a(\mathbf{r}) + U_{a a} n_a + U_{a b} \tilde{n}_b  \right)\phi_a   + \kappa \phi_b , \nonumber \\
\frac{d \phi_b}{dt} = -\frac{i}{\hbar}\left(-\frac{\hbar^2}{2 m}\nabla^2+V_b(\mathbf{r}) + U_{b b} n_b + U_{a b} \tilde{n}_a   \right)\phi_b   + \kappa^* \phi_a,
  \label{eqStochasticEquations}
\end{eqnarray}
where $n_j=\left|\phi_{j}\right|^2 -\frac{1}{dV}$, $\tilde{n}_j=\left|\phi_{j}\right|^2 -\frac{1}{2 dV}$, $dV$ is the volume element of the grid on which the simulation is carried out, and the terms proportional to $1/dV$ correspond to vacuum corrections. It is a peculiarity of the Wigner representation that, up to the vacuum correction terms, Eqs.~(\ref{eqStochasticEquations}) look identical to those of the coupled semi-classical nonlinear Schr{\"{o}}dinger equation. The quantum statistical nature of the Wigner equations enters due to the fact that Eqs.~(\ref{eqStochasticEquations}) are run many times and stochastically averaged, with each run using different set of random initial conditions describing (in this case) the noise on a coherent state.

We integrate Eqs.~(\ref{eqStochasticEquations}) numerically in one, two and three spatial dimensions using the numerical package XMDS2 \cite{dennis2013}.  For the one- and two-dimensional simulations, the dimensional reduction is achieved by estimating the mode shape in three dimensions, and using dimensionally reduced values for $U_{ij}$ that match the zero-dimensional reduction given by Eq.~(\ref{eqChiUequivalence}).  To make relevant comparisons between equivalent situations with different numbers of spatial dimensions, this mode shape is typically chosen to match the chemical potentials of the initial states.  

\section{Effects of trap geometries on squeezing in 1D} \label{sec:1Dsim}

A one-dimensional simulation is sufficient to examine the hypothesis that the squeezing will be higher for a BEC where the spatial mode has a more constant density profile.  We compare squeezing for fields starting in the ground states of a harmonic trap with negligible nonlinearity (a Gaussian), a harmonic trap with strong nonlinearity (Thomas-Fermi), and a constant potential (constant density). Using Eq.~(\ref{eqChiUequivalence}) we find the following equivalences
\begin{align}
\hbar \chi &= U \sqrt{\frac{m^3 \omega_x \omega_y \omega_z}{8 \pi^3 \hbar^3}}, &&({\textrm{Gaussian}}) \\
\hbar \chi &= \frac{4}{7} \left( \frac{15 U \omega_x \omega_y \omega_z}{16 \pi \sqrt{2}} \right)^{2/5} \left( \frac{m}{N} \right)^{3/5}, &&({\textrm{Thomas-Fermi}}) \\
\hbar \chi &= \frac{U}{V}, &&{\textrm{(constant density)}}
\end{align}
where $\omega_x$, $\omega_y$, $\omega_z$ are the angular frequencies of the harmonic trap, and $N$ is the number of particles in the mode of interest.

\begin{figure}
    \centering
    \includegraphics[width=8cm]{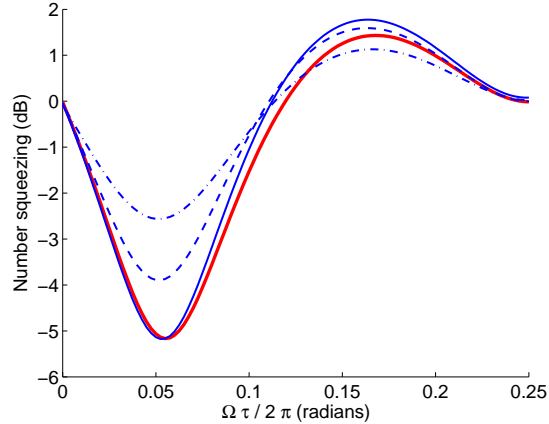}
    \caption{Effects of mode shape on squeezing. The plot shows the normalised number variance in mode $|a \rangle$ as a function of the time the final coupling pulse is applied. We compare the analytic two-mode solution (thick red), the constant density mode (solid blue), the Thomas-Fermi mode (dashed) and the Gaussian mode (dash-dotted). The nonlinearities of the various 1D simulations have been adjusted for their mode shape so that they are equivalent to that of the 0D, two-mode mode nonlinearity. Parameters used were $n_a = n_b =1 \times 10^5$, $\chi_{aa}=0.03 \text{s}^{-1}, \chi_{ab}=\chi_{bb}=0, \tau_{\mathrm{hold}}=3\times 10^{-4}$s, $\phi=4.0$. Maximum squeezing is achieved when the atomic density profile is the most uniform.}
    \label{figModalSqueezingEffects1D}
\end{figure}

Choosing parameters such that the squeezing occurs fast enough that the mode shape is reasonably stable over the simulation gives the best squeezing, and allows us to make a fair comparison between the squeezing achievable by different mode shapes without being concerned that different modes will change shapes at different rates, ruining mode-matching. The results of this simulation are shown in Figure~\ref{figModalSqueezingEffects1D}, which clearly confirms the hypothesis that maximum squeezing is obtained when the mode shape is closest to constant density.

One way to avoid any requirement to engineer and maintain specific atomic modes is to post-select a spatial mode.  Rather than considering the variance in the total number of atoms in each internal state, it may be possible to consider only the statistics in a spatially filtered area of the atomic cloud.  For example, suppose the atoms were in a Gaussian atomic distribution, but we consider a mode that only includes the atoms within $p$ standard deviations of the centre. For low $p$, this mode shape is considerably closer to the constant density case than a standard Gaussian. The equivalence between the single mode case and this mode is given by
\begin{equation}
\hbar\chi = U \frac{\erf (\sqrt{2p})^3}{\erf(p)^6} \sqrt{\frac{m^3 \omega_x \omega_y \omega_z}{8 \pi^3 \hbar^3}}.
\end{equation}
Such spatial filtering can easily produce higher squeezing, but there is the obvious disadvantage that it  significantly reduces the effective number of atoms.  This is a serious issue for most applications of squeezed sources, such as interferometry, where the signal to noise scales with the square root of the flux.  We will focus on methods for producing squeezing with large atomic number.

\section{Using a $\pi$-pulse to improve mode-matching} \label{sec:pipulse}

Modal mismatch is generated by the different atomic potentials seen by the two internal states.  A straightforward method of minimising this effect is to apply a $\pi$-pulse to swap the populations of the internal states after half of the evolution.  As can be seen by Eqs.~\eqref{eqa1} and \eqref{eqb1}, such a pulse also leaves the quantum statistics of each state unchanged.  The idea of this scheme is to make the effective potential seen by the atoms in state $\ket{a}$ during the second half of the hold time very similar to that seen by the atoms that were in state $\ket{a}$ during the first half of the hold time, and will only work if the modes $\ket{a}$ and $\ket{b}$ have an equal occupation at the start of the hold time, or if $\chi_+=\chi_{aa} + \chi_{bb} - 2\chi_{ab}=0$.  In the latter case, this reduces to a spin-echo pulse, and as the squeezing effects of the $\Delta\omega$ term in Eq.~\eqref{eqNonlinearJosephson} are cancelled in this case, there will be no squeezing.  However, in the limit of short hold times, for equal occupations of the two modes, and when $\chi_+\neq 0$, this technique reduces the difference in the mode shapes after the hold time.  In practice this will never work perfectly, as $\psi_a(\mathbf{r})$ and $\psi_b(\mathbf{r})$ will typically have evolved by the time the $\pi$-pulse is applied, and so the spatial dynamics in the second half of the hold time will not be quite the same as those during the first half.  Nonetheless, in many regimes, such a $\pi$-pulse can produce some improvement.  More importantly, as this scheme accepts that the two modes will change shape, and attempts to make them change in a similar fashion, it allows the possibility of much longer hold times. That is, if we are no longer constrained to hold times short enough such that the mode shapes of the atoms in states $|a \rangle$ and $|b \rangle$ do not change significantly, much more squeezing can be extracted from the system.

\begin{figure}
    \centering
    \includegraphics[width=8cm]{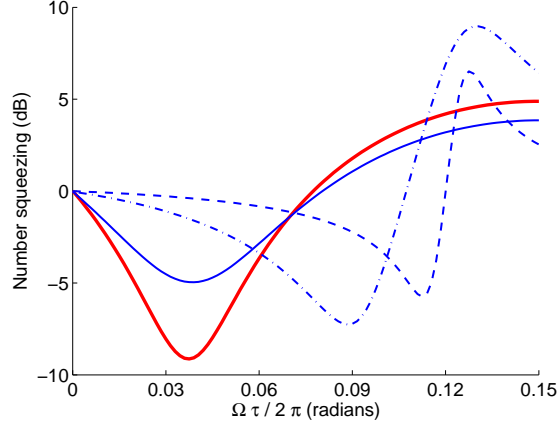}
    \caption{Effect of a $\pi$-pulse halfway through the hold time. Common parameters used were $n_a = n_b =2 \times 10^5$, $\chi_{aa}=0.03\,\text{s}^{-1}$, $\chi_{ab}=\chi_{bb}=0$. Solid red line: Analytic two-mode solution, $\tau_{\mathrm{hold}}=3\times 10^{-4}$s, $\phi=3.1$ (10.9dB squeezing); Solid blue line: no $\pi$-pulse, $\tau_{\mathrm{hold}}=3\times 10^{-4}$s, $\phi=3.1$ (4.9dB squeezing);  Dashed blue line: with $\pi$-pulse, $\tau_{\mathrm{hold}}=3\times 10^{-4}$s, $\phi=3.1$ (5.8dB squeezing); Dash-dotted blue line: with $\pi$-pulse and the longer hold time $\tau_{\mathrm{hold}}=5\times 10^{-4}$s allowed by the application of the pulse, $\phi=3.87$ (7.2dB squeezing).}
    \label{piPulseFig}
\end{figure}

These effects are illustrated in Fig.~\ref{piPulseFig}, which shows the best squeezing obtainable from a specific system with and without the $\pi$-pulse, as well as the extra squeezing that can be obtained by allowing a longer hold time as well as the $\pi$-pulse.

\section{Squeezing in 2D and 3D} \label{sec:2D3Dsqueezing}

\begin{figure}
  \centering
  \begin{subfigure}{.5\textwidth}
    \centering
    \includegraphics[width=6cm]{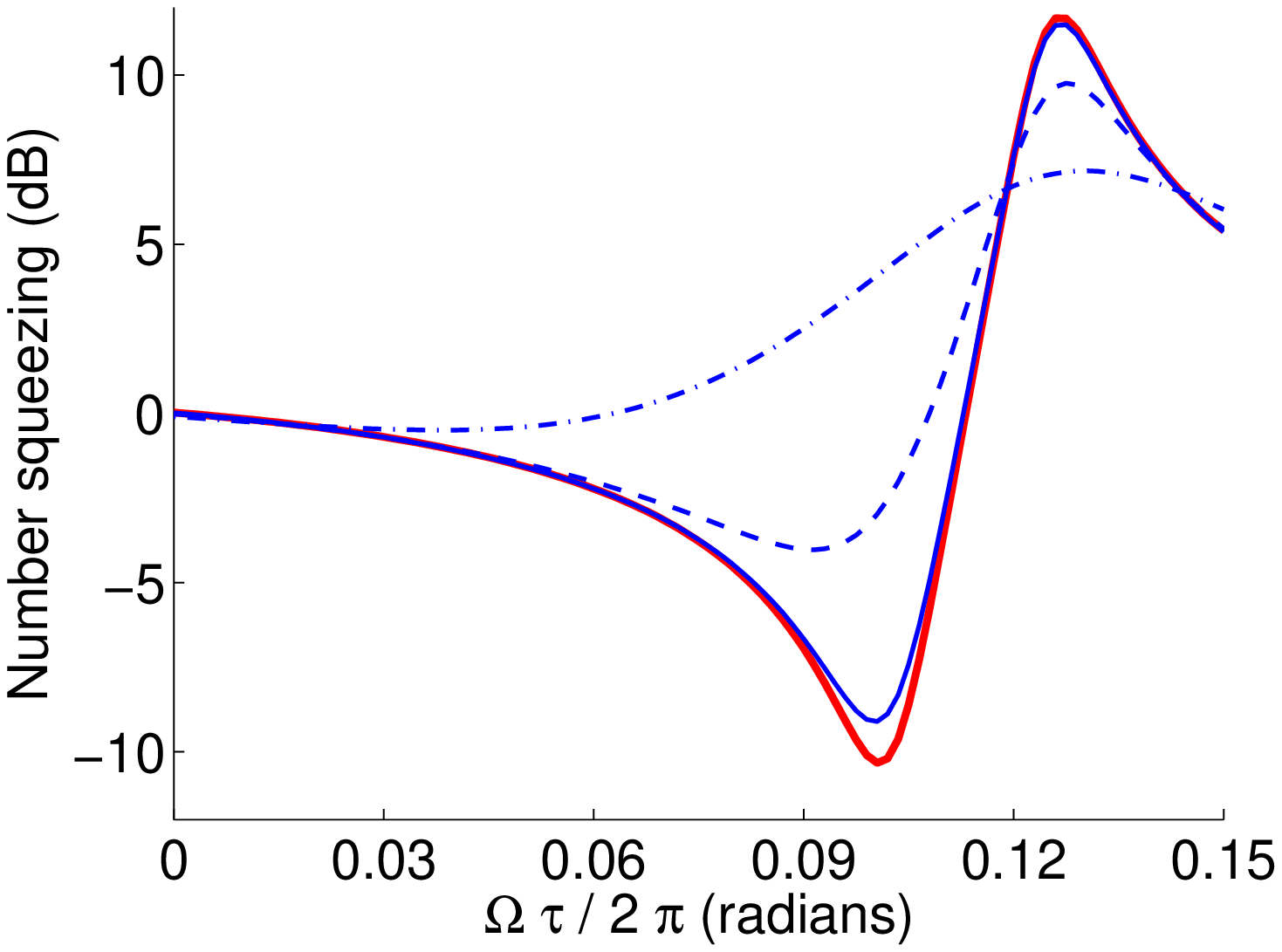}
    \label{figDimensionalSqueezingEffects:sub1}
  \end{subfigure}%
  \begin{subfigure}{.5\textwidth}
    \centering
    \includegraphics[width=6cm]{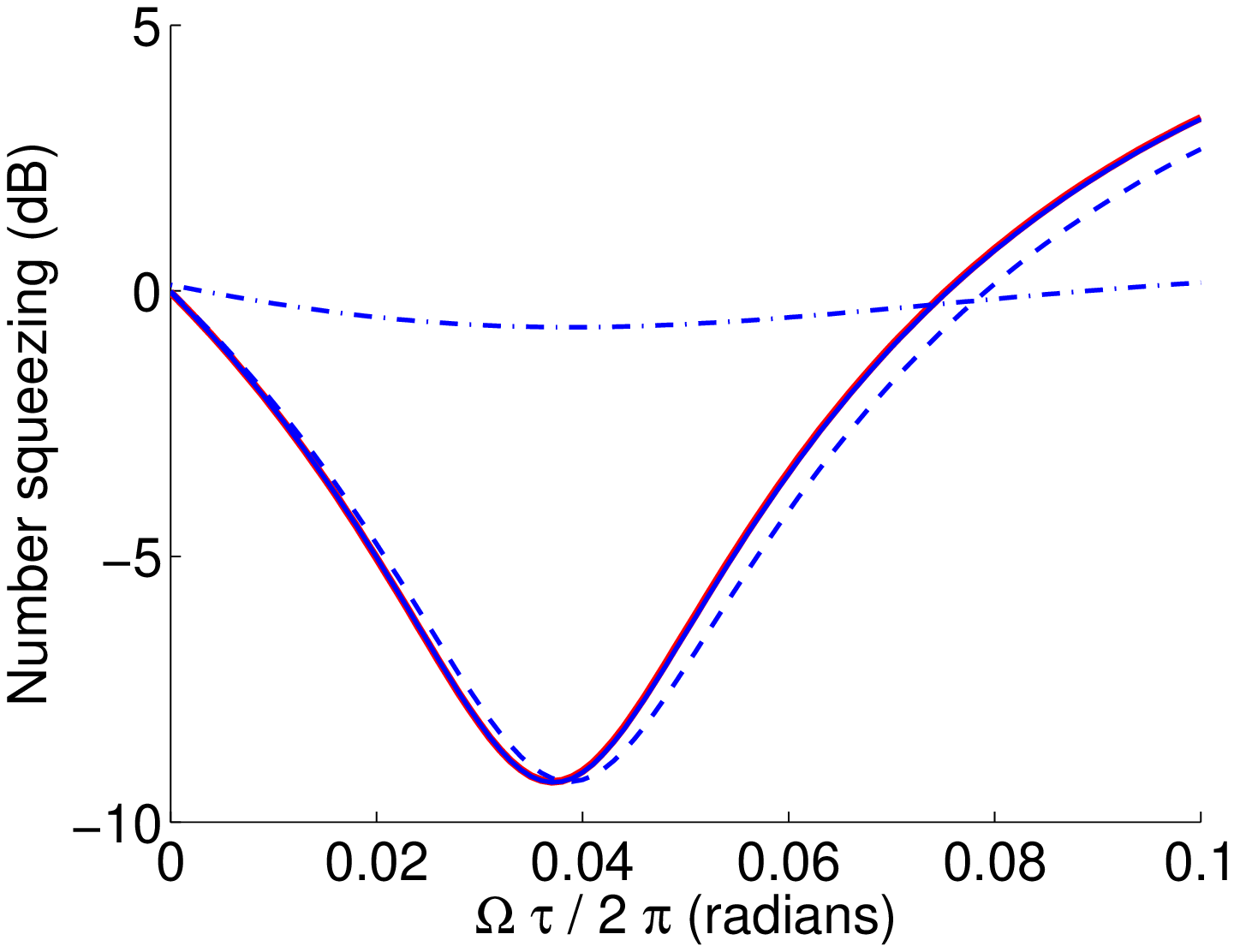}
    \label{figDimensionalSqueezingEffects:sub2}
  \end{subfigure}
\caption{Effects of dimension on squeezing in two different nonlinear regimes, assuming that the chemical potential is constant for all simulations. If two nonlinearities are non-zero, the deleterious effects of higher dimensions on squeezing are exacerbated. Thick red line, analytic two-mode solution; solid blue line, 1D; dashed line, 2D; dash-dotted line, 3D. (a) Parameters: $n_a = n_b =2 \times 10^5$, $\chi_{aa}=\chi_{bb}=0.03\,\text{s}^{-1}$, $\chi_{ab}=0$, $\tau_{\mathrm{hold}}=3\times 10^{-4}$s, $\phi=6.2$. (b) $n_a = n_b =2 \times 10^5$, $\chi_{aa}=0.03\,\text{s}^{-1}$, $\chi_{bb}=\chi_{ab}=0$, $\tau_{\mathrm{hold}}=3\times 10^{-4}$s, $\phi=3.1$.}
  \label{figDimensionalSqueezingEffects}
\end{figure}

The simulations in the previous two sections were one dimensional, and show how having spatially constant density and using a $\pi$-pulse will improve the resultant squeezing.  While these conclusions remain valid for simulations in two or three dimensions, adding each additional dimension increases the number of spatial modes in a given energy range.  For BECs of different dimension at the same energy per particle, this leads to a degradation of squeezing, even when the potential (and the resulting atomic density) is constant across the bulk of the trap.  This is demonstrated in Fig.~\ref{figDimensionalSqueezingEffects}, where we show the squeezing generated in 1D, 2D and 3D BECs for two different parameter regimes. In both cases we take the BEC to be trapped in a box potential, which in the Thomas-Fermi limit leads to a constant density profile across the BEC. In Figure~\ref{figDimensionalSqueezingEffects}a we consider the case where $U_{aa}=U_{bb}\neq 0$, $U_{ab}=0$, while in Figure~\ref{figDimensionalSqueezingEffects}b we consider the case $U_{aa}\neq 0$, $U_{ab}=U_{bb}=0$. In both cases it is clear that number squeezing is best in lower dimensions, and becomes degraded or non-existent as we move to a three dimensional system; the effect is worse if more than one nonlinearity is non-zero.

To find the trap geometries that produce the best squeezing, it is important to understand the root cause of this loss of squeezing in higher dimensional traps.  The dependence on dimensionality is not explained by the processes described in section \ref{sec:MMdescription}, which focus on the effects of inhomogeneity within the trap.  Fundamentally, this degradation of the squeezing is due to nonlinearity-induced coupling between different momentum modes.  

We can understand the origin and scaling of this degradation of squeezing by considering the response of the condensate to small fluctuations about the mean field.  We do this using Bogoliubov theory \cite{PethickSmith}, which considers the first order quantum-mechanical fluctuations about the mean field.  To simplify the analysis we consider the case in which $U_{aa} \neq 0$, $U_{bb}=U_{ab} = 0$ and the condensate has a spatially-constant density in a box of side lengths $L_x$, $L_y$ and $L_z$.  In this limit the occupation of the non-zero momentum modes (the non-condensed fraction) in the $\ket{a}$ internal state at the end of the pulse sequence is (a derivation is given in \ref{appendixBogDerivation})
\begin{align}
  n_a(\mathbf{k}, t_3) &= \expect{\hat{a}^\dagger(\mathbf{k},t) \hat{a}(\mathbf{k},t)} = \left[n_a \chi_{aa} \tau_\text{hold} \cos(\theta) \sinc\left(\omega_\mathbf{k} \tau_\text{hold}\right) \right]^2, \label{eqBogOccupation}
\end{align}
where the Bogoliubov mode frequency is $\omega_\mathbf{k} = \sqrt{\omega^0_\mathbf{k}(\omega^0_\mathbf{k} + 2 \chi_{aa} n_a)}$, and the free particle frequency is $\omega^0_\mathbf{k} = \hbar \mathbf{k}^2/2 M$.  The non-zero momentum modes can be expected to have an impact on the squeezing of the system when their occupation is a non-negligible fraction of the total number of atoms.  The fractional occupation of these modes is
\begin{align}
  \frac{N_{\mathbf{k}\neq 0}(t)}{N} &= \frac{1}{N}\sum_{\mathbf{k}\neq 0} n_a(\mathbf{k}, t), \label{eqCondensateDepletion}
\end{align}
where the sum over $\mathbf{k} \neq 0$ is taken over the available non-zero momentum modes $\mathbf{k} = (2\pi/L_x) n_x \hat{\mathbf{x}} + (2\pi/L_y) n_y \hat{\mathbf{y}} + (2\pi/L_z) n_z \hat{\mathbf{z}}$ with $n_x$, $n_y$ and $n_z$ arbitrary integers that are not all zero.

\begin{figure}
    \centering
    \includegraphics[width=10cm]{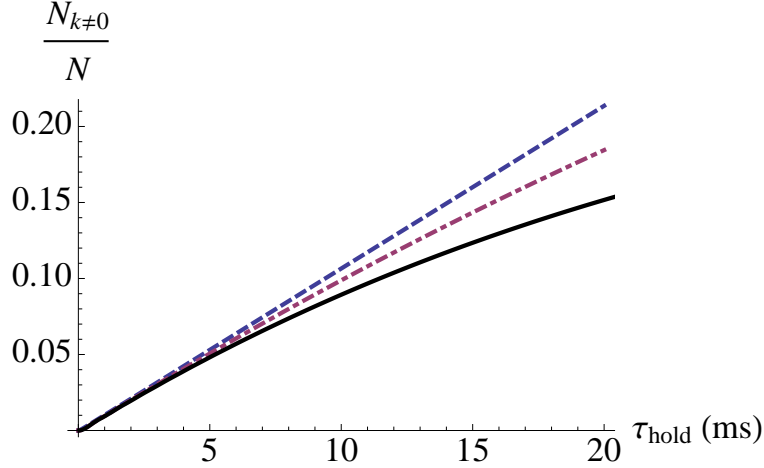}
\caption{Plot of the non-condensed fraction $N_{\mathbf{k}\neq0}/N$ as a function of the hold time $\tau_\text{hold}$.  The blue dashed line is the Bogoliubov integral approximation given in Eq.~(\ref{eqBogFraction1D}), the purple dot-dashed line is the Bogoliubov summation expression given in Eq.~(\ref{eqCondensateDepletion}), and the solid black line is the result of a truncated Wigner simulation.  Parameters: $L=600\,\mu\text{m}$, $N=4 \times 10^5$, $n_a=n_b=2\times 10^5$, $\chi_{aa} = 2.67 \times 10^{-2}\, \text{s}^{-1}$, $\chi_{ab}=\chi_{bb}=0$, $\theta=0$.}
  \label{figBogSqueezingValidation}
\end{figure}

Figure~\ref{figBogSqueezingValidation} demonstrates the agreement between truncated Wigner simulations and the Bogoliubov theory for the non-condensed fraction for parameters where the occupation of these modes approach 15\% of the total occupation of the system.  The disagreement for larger non-condensed fractions occurs because the Bogoliubov theory used in deriving \eqref{eqCondensateDepletion} is not number-conserving as it neglects the effect of the non-zero momentum modes on the condensate.  


To minimise the damaging effects of the non-condensed fraction it is better to operate in a regime where the total occupation of the non-zero momentum modes $N_{\mathbf{k}\neq 0}$ is small, which is also fortuitously the regime in which the Bogoliubov theory is a good approximation, so we can investigate in more detail.  In 1D Eq.~\eqref{eqCondensateDepletion} is
\begin{align}
  \frac{N_{\mathbf{k}\neq 0}(t)}{N} &= \frac{(n_a \chi_{aa} \tau_\text{hold} \cos\theta)^2}{N} \sum_{k \neq 0} \sinc^2\left(\sqrt{\omega^0_{k}(\omega^0_{k}+2 \chi_{aa}n_a)}\tau_\text{hold}\right).
\end{align}
We approximate this sum by the integral
\begin{align}
  \frac{N_{\mathbf{k}\neq 0}(t)}{N} &\approx \frac{(n_a \chi_{aa} \tau_\text{hold} \cos\theta)^2}{N} \int_{-\infty}^{+\infty} \sinc^2 \left(\sqrt{\omega^0_{k}(\omega^0_{k}+2 \chi_{aa}n_a)}\tau_\text{hold}\right) \frac{d k}{\Delta k}, \label{eqCondensateDepletionFirstIntegralApproximation}
\end{align}
where $\Delta k = 2\pi/L$ and $L$ is the length of the 1D box condensate.  This approximation will be valid in the limit that $\omega_{k_\text{min}} \tau_\text{hold} \ll 1$, where $k_\text{min} = 2\pi/L$ as for small $k$ the allowed values $k = n_x 2\pi /L$ will closely sample the region $\xi = \sqrt{\omega_k^0(\omega_k^0 + 2 \chi_{aa} n_a)} \tau_\text{hold} < \pi$ where $\sinc^2(\xi)$ is significant.

Changing integration variables to $\Phi = \omega^{0}_{\mathbf{k}}\tau_\text{hold}$ and defining $\Lambda_a = n_a \chi_{aa} \tau_\text{hold}$, the non-condensed fraction \eqref{eqCondensateDepletionFirstIntegralApproximation} can be written as
\begin{align}
  \frac{N^\text{1D}_{\mathbf{k}\neq 0}(t)}{N} &\approx \frac{1}{N} \frac{\Lambda_a^2 L \cos^2\theta } {\sqrt{2}\pi\sqrt{\frac{\hbar}{M}\tau_\text{hold}}} f^\text{1D}(\Lambda_a), \label{eqBogFraction1D}
\end{align}
where
\begin{align}
  f^\text{1D}(\Lambda) &= \int_0^{\infty} \frac{1}{\sqrt{\Phi}}\sinc^2\sqrt{\Phi(\Phi + 2 \Lambda)}\, d\Phi.
\end{align}
In 2D and 3D the equivalent expressions are
\begin{align}
  \frac{N^{\text{2D}}_{\mathbf{k}\neq 0}(t)}{N} &\approx \frac{1}{N} \frac{\Lambda_a^2 A \cos^2\theta } {4\frac{\hbar}{M}\tau_\text{hold}} f^\text{2D}(\Lambda_a),  \label{eqBogFraction2D}\\
  \frac{N^{\text{3D}}_{\mathbf{k}\neq 0}(t)}{N} &\approx \frac{1}{N} \frac{\Lambda_a^2 V \cos^2\theta } {\sqrt{2}\pi^2\left(\frac{\hbar}{M}\tau_\text{hold}\right)^{3/2}} f^\text{3D}(\Lambda_a), \label{eqBogFraction3D}
\end{align}
where
\begin{align}
  f^\text{2D}(\Lambda) &= \int_0^{\infty} \sinc^2\sqrt{\Phi(\Phi + 2 \Lambda)}\, d\Phi,\\
  f^\text{3D}(\Lambda) &= \int_0^{\infty} \sqrt{\Phi} \sinc^2\sqrt{\Phi(\Phi + 2 \Lambda)}\, d\Phi,
\end{align}
and $A$ is the area of the two-dimensional system and $V$ is the volume of the three-dimensional system.  Figure~\ref{figBogSqueezingValidation} demonstrates the agreement in 1D between the integral approximation to $N_{\mathbf{k}\neq 0}/N$ in \eqref{eqBogFraction1D}, the summation expression in \eqref{eqCondensateDepletion} and a truncated Wigner simulation.

We wish to examine the scaling of the non-condensed fraction $N_{\mathbf{k}\neq0}/N$ with the physical size of the condensate.  Other system parameters such as the total number $N$ and the occupation of the $\ket{a}$ and $\ket{b}$ states at $t_1$ ($n_a$ and $n_b$, respectively) are kept constant and chosen to maximise the squeezing of the equivalent two-mode model presented in Section~\ref{secTwoModeAnalytic}.  In particular, this fixes the values of $\theta$, $\phi$, and $\lambda_{ij} = \chi_{ij}\tau_\text{hold}$.  As the size of the condensate is scaled, $\hbar \chi_{ij} = U^\text{1D}_{ij}/L = U^\text{2D}_{ij}/A = U^\text{3D}_{ij}/V$ will change, but we will vary $\tau_\text{hold}$ to keep $\lambda_{ij}$ constant.  Therefore $\Lambda_a = n_a\chi_{aa}\tau_\text{hold}$ will also be constant.  With this in mind, we can write $\tau_\text{hold}$ in terms of the system size and the nonlinearity $U_{aa}$.  The non-condensed fraction can then be shown to scale with the physical size of the condensate as:
\begin{align}
  \frac{N^\text{1D}_{\mathbf{k}\neq 0}(t_3)}{N} &\approx \sqrt{L} \cdot \frac{1}{N} \sqrt{\frac{M U^\text{1D}_{aa} n_a}{2 \hbar^2 \pi^2}} \Lambda_a^{3/2} \cos^2\theta f^\text{1D}(\Lambda_a) &&\propto \sqrt{L}, \\
  \frac{N^\text{2D}_{\mathbf{k}\neq 0}(t_3)}{N} &\approx 1 \cdot \frac{1}{N} \frac{M U^\text{2D}_{aa} n_a }{4 \hbar^2} \Lambda_a \cos^2\theta f^\text{2D}(\Lambda_a) &&\propto \text{constant}, \\
  \frac{N^\text{3D}_{\mathbf{k}\neq 0}(t_3)}{N} &\approx \frac{1}{\sqrt{V}} \cdot \frac{1}{N} \left(\frac{M U^\text{3D}_{aa} n_a}{\hbar^2}\right)^{3/2} \frac{\Lambda_a^{1/2} \cos^2\theta f^\text{3D}(\Lambda_a)}{\sqrt{2} \pi^2} &&\propto \frac{1}{\sqrt{V}}.
\end{align}
The non-condensed fraction $N_{\mathbf{k}\neq 0}/N$ thus scales differently with trap size in each of 1D, 2D and 3D.  For one-dimensional traps, the non-condensed fraction increases with system size, whereas for 2D traps it is constant.  For three-dimensional traps, the non-condensed fraction \emph{decreases} with volume.  This suggests that the best system to use will be a three-dimensional box potential, and the system size can be increased until the condensate depletion becomes negligibly small.

\begin{figure}
  \centering
  \begin{subfigure}{.48\textwidth}
    \centering
    \includegraphics[width=\textwidth]{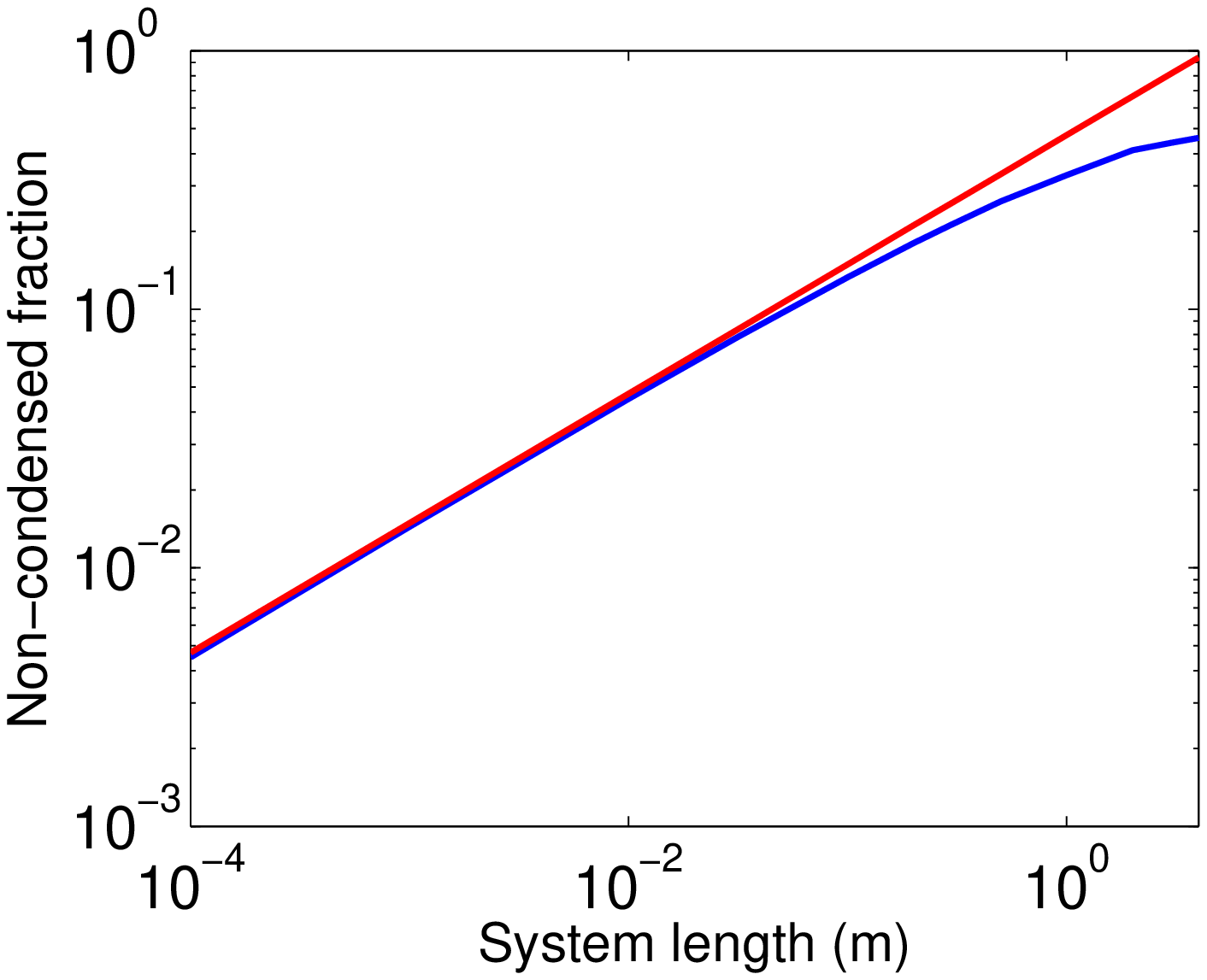}
  \end{subfigure}
  \begin{subfigure}{.48\textwidth}
    \centering
    \includegraphics[width=\textwidth]{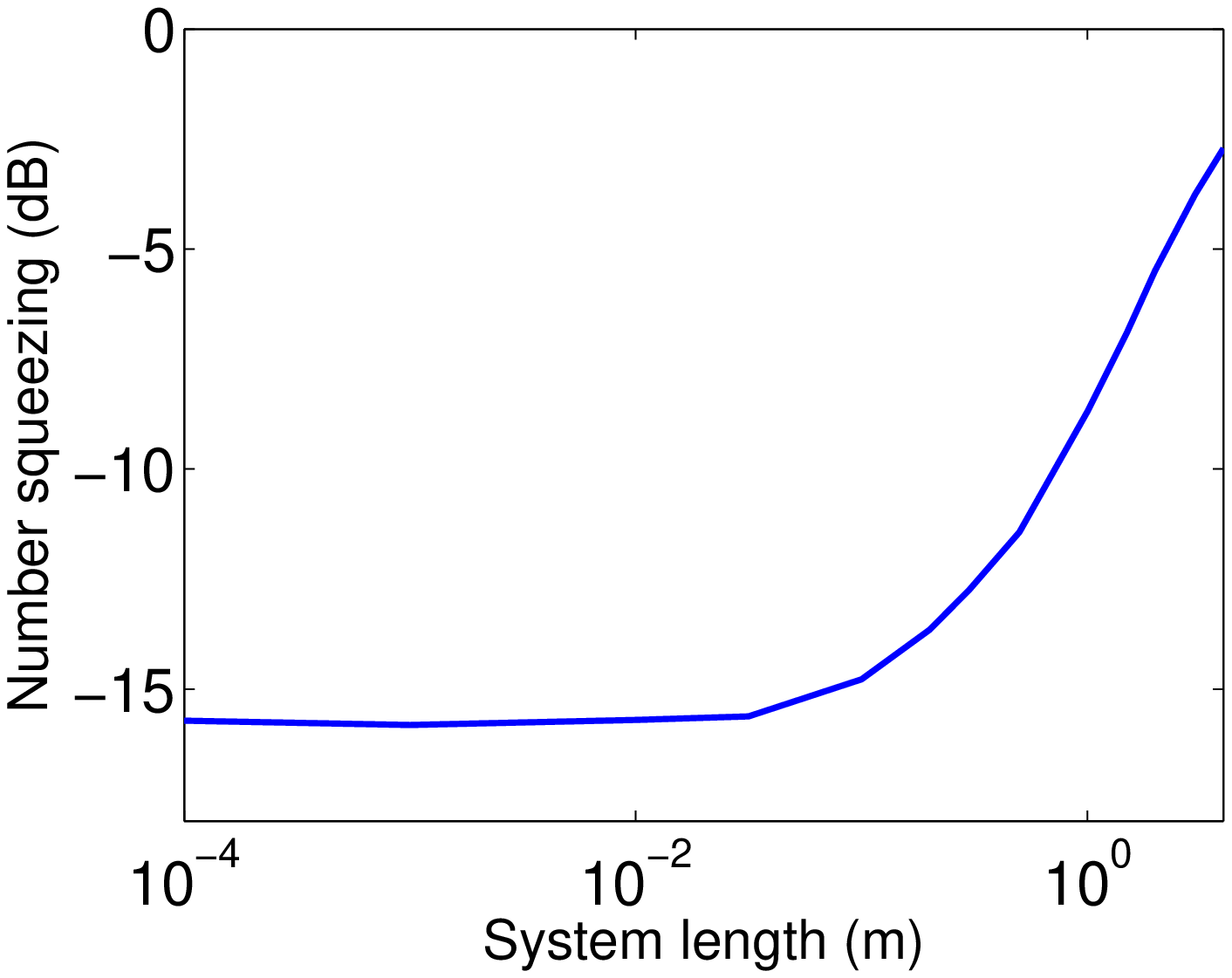}
  \end{subfigure}
\caption{Non-condensed fraction and squeezing as function of trap size $L$ in a 1D trap. (Left) The proportion of the non-condensed fraction in state $|a\rangle$ relative to total particle number, with the Bogoliubov theory model in red and the full numerical solution in blue. (Right) Relative number squeezing in mode $|a\rangle$ as a function of trap size. Parameters: $N=1\times 10^5$, $U_{aa} = 8.9\times 10^{-40}$Jm, $U_{ab}=U_{bb}=0$, $\tau_{\mathrm{hold}} = 10L$\,s.}
  \label{fig1D_squeezing_and_bog_scaling}
\end{figure}

\begin{figure}
  \centering
  \begin{subfigure}{.48\textwidth}
    \centering
    \includegraphics[width=\textwidth]{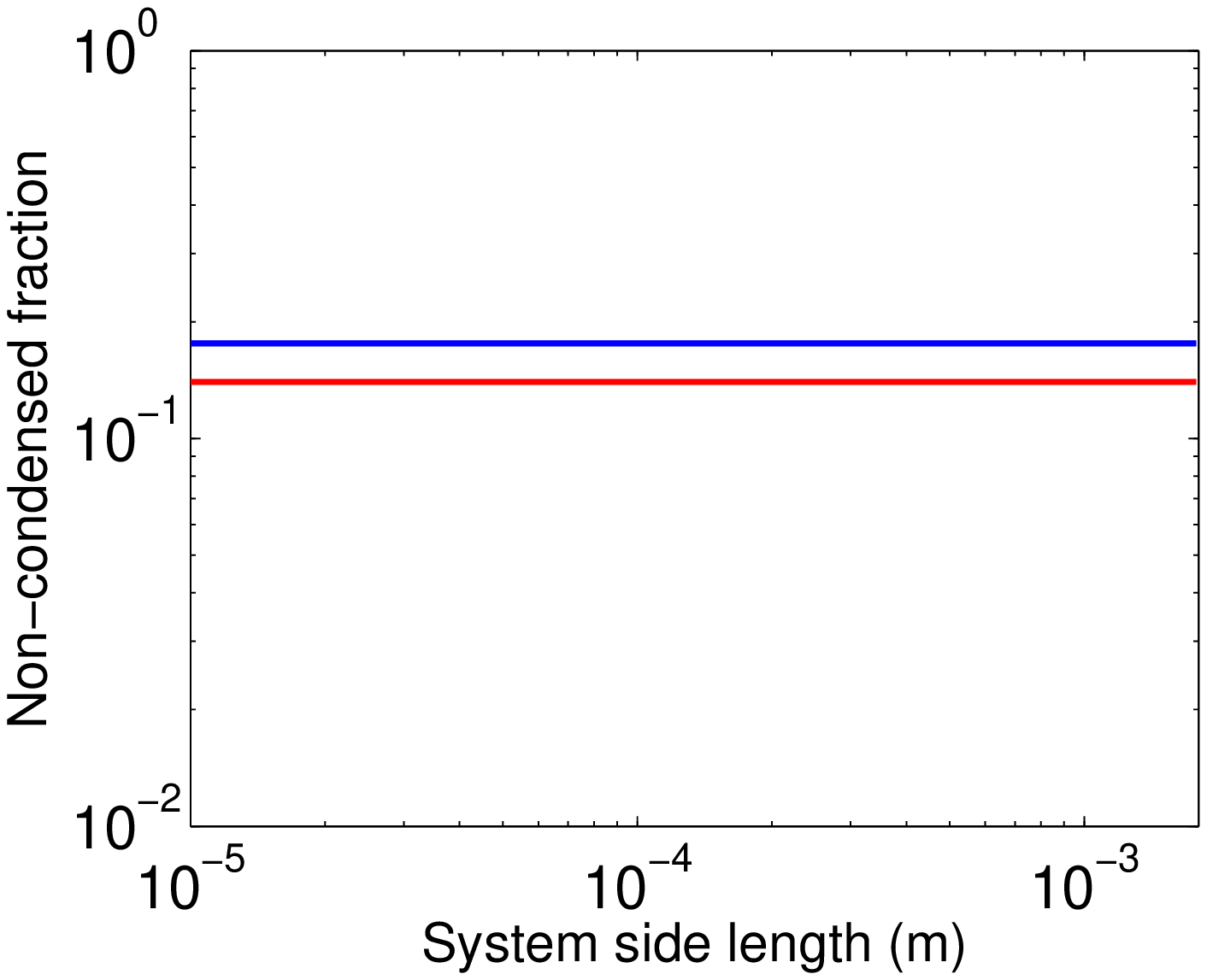}
  \end{subfigure}
  \begin{subfigure}{.48\textwidth}
    \centering
    \includegraphics[width=\textwidth]{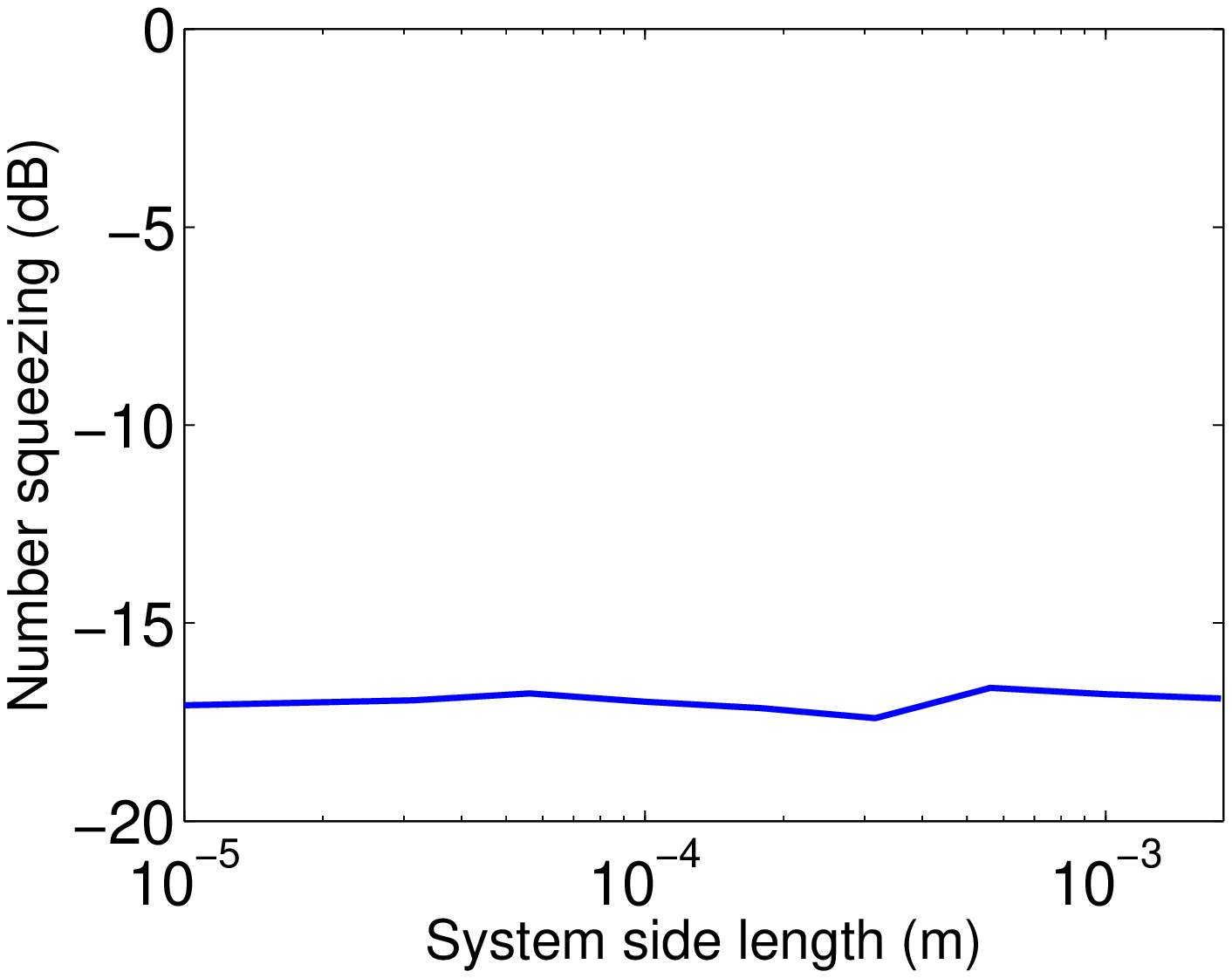}
  \end{subfigure}
\caption{Non-condensed fraction and squeezing as function of trap size $L$ in a 2D trap. (Left) The proportion the non-condensed fraction in state $|a\rangle$ relative to total particle number, with the Bogoliubov theory model in red and the full numerical solution in blue. (Right) Relative number squeezing in mode $|a\rangle$ as a function of trap size. Parameters: $N=1\times 10^5$, $U_{aa} = 8.9\times 10^{-44}$Jm$^2$, $U_{ab}=U_{bb}=0$, $\tau_{\mathrm{hold}} = 10^{5} L^2$\,s.}
  \label{fig2D_squeezing_and_bog_scaling}
\end{figure}

\begin{figure}
  \centering
   \includegraphics[width=7cm]{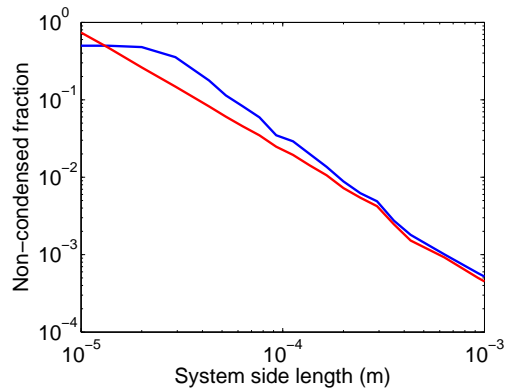}
\caption{Non-condensed fraction and squeezing as function of trap size $L$ in a 3D trap. (Left) The proportion the non-condensed fraction in state $|a\rangle$ relative to total particle number, with the Bogoliubov theory model in red and the full numerical solution in blue. (Right) Relative number squeezing in mode $|a\rangle$ as a function of trap size. Parameters: $N=1\times 10^4$, $U_{aa} = 4.7\times 10^{-49}$Jm$^3$, $U_{ab}=U_{bb}=0$, $\tau_{\mathrm{hold}} = 1.25 \times 10^{11} L^3$\,s.}
 \label{fig3D_Bog_mode_occupation_scaling}
\end{figure}

Figures~\ref{fig1D_squeezing_and_bog_scaling}--\ref{fig3D_Bog_mode_occupation_scaling} show the non-condensed fraction predicted by the Bogoliubov analysis compared to that predicted by a full numerical simulation in one, two and three dimensions, as a function of the trap size. Also shown in the 1D and 2D cases is the maximum amount of number squeezing obtainable, as predicted by the numerical simulation. The full squeezing scaling curve is not shown in 3D, as for the smaller trap sizes the TWA method breaks down, and results are unreliable. For the larger trap sizes with small non-condensate fractions, however, the TWA simulation remained valid, and we obtained relative number squeezing of 12.7 dB. The traps are assumed to have a constant potential, which minimises the effects of the multimode issues described in Section \ref{sec:MMdescription}.  We see that the squeezing does indeed degrade at large trap sizes in 1D, but remains constant in 2D. For small trap sizes, the squeezing in 1D asymptotically approaches that of the ideal 0D system.

We note that the non-smooth nature of the squeezing prediction in 2D is due to the stochastic nature of the simulations, and the necessity of averaging over fewer quantum trajectories due to the very high number grid points in this regime. The non-smooth nature of the non-condensate mode fraction in 3D, on the other hand, is due to the number of modes being both discrete and low in this regime, and counting only the non-condensed fraction in modes with an energy less than than $\mu$, the chemical potential.

The simulation shown in Fig.~\ref{fig3D_Bog_mode_occupation_scaling} demonstrates that provided particle number is kept fixed, the number of accessible states for three-dimensional traps decreases with volume in 3D.  This is because the energy of the ground state of the BEC (the chemical potential, $\mu$) is reducing faster than the density of states at low energies is increasing, which in turn results in fewer states that can become dynamically occupied.  This scaling of the estimated Bogoliubov fraction in 3D suggests that strong squeezing will be found in very large, weakly trapped condensates, provided the trapping potentials are box-like. Such box-like traps have already been demonstrated and shown to be capable of producing BECs \cite{gaunt2013}.

Although systematic analysis of squeezing across a wide range of large 3D traps and interaction strengths is impossible due to the breakdown of the TWA method, fortunately the method is accurate when the highest squeezing is produced.  While inter-state scattering lengths of most atomic species and states are not well characterised, extremely high squeezing can be found in systems with large, three-dimensional box potentials, and asymmetry in the three scattering lengths.

As a demonstration of the plausibility of such a scheme, we carried out a full 3D simulation using the geometry of the optical box potential described in Ref.~\cite{gaunt2013}, using the same potentials (a cylindrical ``box'' with length 63\,$\mu$m and diameter 30\,$\mu$m) and condensate number ($N=10^5$ $^{87}$Rb atoms). We set one s-wave scattering length to $5$\,nm and the others to zero: $a_{00}=5$\,nm, $a_{10}=a_{11}=0$. The associated nonlinearities are given by Eq.~(\ref{eqUij}).  This system produced number squeezing of 15dB, and an extremely low non-condensed fraction of $4 \times 10^{-4}$. Both of these numbers are in agreement with the Bogoliubov calculations.

\section{Conclusion}
\label{sectionConclusion}

The large Kerr-type nonlinearities present between neutral atoms in BEC present the possibility of significant squeezing, but they also cause potentially complicated multimode behaviour when large BECs are used.  Furthermore, the nonlinearities between multiple internal atomic states are not yet known for many atomic species, and often have sensitive dependence on external magnetic fields due to Feshbach resonances \cite{cornish:2000,altin:2010}.  This paper analyses the requirements for producing useful squeezing via these nonlinearities for systems with large numbers of atoms.  

We have presented an analytic solution to the two-mode model of atomic fields with a strong Kerr nonlinearity, allowing for a rapid estimate of the atomic number difference squeezing that can be obtained from these systems as scattering length data is obtained. In general, it predicts arbitrarily good squeezing in the two-mode limit. We then examined trap geometries in which this limit can be approached.

Using a constant potential in the trap (a `box' trap) allows the BEC density to be spatially uniform throughout the process.  This is shown to alleviate the difficulties with choosing correct hold times and phases for the coupling pulses, and also alleviates the majority of the mode-matching issues for the final recombination of the two modes. If such a box potential is not available, the application of a $\pi$-pulse part way through the hold period can also improve mode matching by imparting the same modal dynamics to both states.

Even in the best case, where a box potential is used, the existence of multiple possible momentum modes allows the possibility of cross-mode scattering via nonlinear processes, which can degrade or totally remove any number squeezing. This degradation is visible only in full 3D simulations, and not lower dimensional approximations, due to its dependence on the number of available modes. It is worth noting that despite their difficulty, in multimode systems with large atom number, full 3D simulations need to be carried out to accurately determine the presence of number or spin squeezing.

Excluding these modes by making them energetically unfavourable can reduce the system to a single mode problem, making our analytic solution valid, but requires extremely tight trapping potentials. Such a regime also strongly limits the number of atoms that can be trapped, partly due to the rate at which the required trapping frequencies grow with atom number, and partly due to inelastic scattering rates.

Using a Bogoliubov analysis, however, it was shown that the coupling between different modes could be minimised in a second regime, namely the \textit{weak} trapping limit for 3D traps. This latter regime is in fact ideally compatible with large atomic numbers, and shows the prospect of extremely high squeezing.  A 3D TWA simulation using an experimentally demonstrated trapping potential yielded $15$\,dB of squeezing with $10^5$ particles. In practice, the only limit to the number of atoms in such traps would be the stability of the trap, as the BEC would become increasingly sensitive to fluctuations as the energy per particle is reduced as the volume increases.

\ack
JH acknowledges the support of an ARC Future Fellowship.

\clearpage

\appendix
\section{Derivation of two mode squeezing formula}
\label{appendixTwoModeDerivation}
In this Appendix we derive the equations for the number variances. We use the model described in Section \ref{secTwoModeAnalytic}, where we consider a two mode system with $N$ particles, initially prepared as a coherent state with all the population in mode $|a\rangle$ at time $t=0$. A coupling pulse is then applied, assumed to be on resonance, so that the system evolves under the Hamiltonian
\begin{equation}
\hat{H_1} = \hbar \Omega (\hat{a}^{\dagger} \hat{b} + \hat{b}^{\dagger} \hat{a} )
\end{equation}
until time $t=t_1$. From time $t_1$ until time $t=t_2$ the coupling is turned off, and the system evolves under the Hamiltonian
\begin{equation}
\hat{H_2} = \hbar\frac{\chi_{aa}}{2} \hat{a}^{\dagger} \hat{a}^{\dagger} \hat{a} \hat{a}
          + \hbar\chi_{ab} \hat{a}^{\dagger} \hat{a} \hat{b}^{\dagger} \hat{b}
          + \hbar\frac{\chi_{bb}}{2} \hat{b}^{\dagger} \hat{b}^{\dagger} \hat{b} \hat{b}.
\end{equation}
Finally, from time $t_2$ until time $t=t_3$, the coupling is once again applied with a phase offset of $\phi$ relative to the first time, and the system evolves via
\begin{equation}
\hat{H_3} = \hbar\Omega (e^{i\phi} \hat{a}^{\dagger} \hat{b} + e^{-i\phi} \hat{b}^{\dagger} \hat{a} ).
\end{equation}

We will work in the Heisenberg picture, and derive expressions for $\hat{a}(t_3)$ and $\hat{b}(t_3)$ in terms of $\hat{a}(t_2)$ and $\hat{b}(t_2)$, which in turn can be expressed in terms of $\hat{a}(t_1)$ and $\hat{b}(t_1)$. We will show that the state of the system at time $t_1$ consists of separate coherent states in modes $|a\rangle$ and $|b\rangle$, enabling us to calculate expectation values at time $t_3$ in terms of the known state at $t_1$.

In order to have more compact expressions we will use the notation
\begin{eqnarray}
\hat{a}(t_j) &=& \hat{a}_j \\
\hat{b}(t_j) &=& \hat{b}_j.
\end{eqnarray}

Beginning with the Hamiltonian $\hat{H}_1$, during the period $(0, t_1)$ mode $\hat{a}$ will evolve as
\begin{equation}
\hat{a}_1 = e^{ i \hat{H}_1 t/ \hbar} \, \hat{a}_0 \, e^{-i \hat{H}_1 t/ \hbar}.
\end{equation}
Utilizing the Hadamard lemma
\begin{equation}
e^{\hat{X}} \hat{Y} e^{-\hat{X}} = \hat{Y} + [\hat{X},\hat{Y}] + \frac{1}{2!}[\hat{X},[\hat{X},\hat{Y}]] + \frac{1}{3!}[\hat{X},[\hat{X},[\hat{X},\hat{Y}]]] + \ldots
\label{eqHadamard}
\end{equation}
we obtain 
\begin{equation}
\hat{a}_1 =  \cos (\Omega t_1) \hat{a}_0 -i \sin (\Omega t_1) \hat{b}_0.
\label{eqa1}
\end{equation}
Similarly, we find
\begin{equation}
\hat{b}_1 = \cos (\Omega t_1) \hat{b}_0 - i \sin (\Omega t_1) \hat{a}_0.
\label{eqb1}
\end{equation}
From the form of (\ref{eqa1}) and (\ref{eqb1}) it is clear that as both modes $\hat{a}$ and $\hat{b}$ began in coherent states, they will remain in coherent states, with only their amplitudes changing. Specifically, at time $t_1$ the system is in a product of coherent states, which we denote $|\alpha, \beta\rangle$ with $\alpha=\sqrt{n_a}$ and $\beta=-i\sqrt{n_b}$ with $n_a + n_b = N$.

Next we consider evolution under $\hat{H}_2$. Taking mode $\hat{a}$ first, we note that as it commutes with the $\hat{b}^{\dagger} \hat{b}^{\dagger} \hat{b} \hat{b}$ term, and since $\hat{a}^{\dagger} \hat{a}^{\dagger} \hat{a} \hat{a}$ commutes with $\hat{a}^{\dagger} \hat{a} \hat{b}^{\dagger} \hat{b}$ we have
\begin{equation}
\hat{a}_2 = e^{i \lambda_{ab} \hat{a}_1^{\dagger} \hat{a}_1 \hat{b}_1^{\dagger} \hat{b}_1 } 
          e^{ i \frac{\lambda_{aa}} {2} \hat{a}_1^{\dagger} \hat{a}_1^{\dagger} \hat{a}_1 \hat{a}_1}\, \hat{a}_1 \,  
          e^{ -i \frac{\lambda_{aa}}{2} \hat{a}_1^{\dagger} \hat {a}_1^{\dagger} \hat{a}_1 \hat{a}_1 }
          e^{-i \lambda_{ab} \hat{a}_1^{\dagger} \hat{a}_1 \hat{b}_1^{\dagger} \hat{b}_1}
\label{eqa2evolution}
\end{equation}
where we have defined $\lambda_{ij} = \chi_{ij} (t_2-t_1)$. Using (\ref{eqHadamard}) to move $\hat{a}_1$ through the exponential, with some algebra one can show that
\begin{eqnarray}
e^{ i \frac{\lambda_{aa}} {2} \hat{a}_1^{\dagger} \hat{a}_1^{\dagger} \hat{a}_1 \hat{a}_1} \hat{a}_1 
         e^{ -i \frac{\lambda_{aa}} {2} \hat{a}_1^{\dagger} \hat{a}_1^{\dagger} \hat{a}_1 \hat{a}_1} &=& \left[ \sum_{n=0} (-i \lambda_{aa})^n (\hat{a}_1^{\dagger} \hat{a}_1)^n \right] \hat{a}_1 \nonumber \nonumber \\
   &=& \exp[-i \lambda_{aa} \hat{a}_1^{\dagger} \hat{a}_1] \hat{a}_1.
\end{eqnarray}
To handle the cross-nonlinearity term in (\ref{eqa2evolution}) we use the identity \cite{louisell}
\begin{equation}
e^{x \hat{a}^{\dagger} \hat{a}} f(\hat{a}, \hat{a}^{\dagger}) e^{-x \hat{a}^{\dagger} \hat{a}} = f(\hat{a}e^{-x}, \hat{a}^{\dagger} e^{x})
\label{eqefeidentity}
\end{equation}
to obtain
\begin{eqnarray}
\hat{a}_2 &=& e^{i \lambda_{ab} \hat{a}_1^{\dagger} \hat{a}_1 \hat{b}_1^{\dagger} \hat{b}_1 } 
          e^{-i \lambda_{aa} \hat{a}_1^{\dagger} \hat{a}_1} \, \hat{a}_1 \,
          e^{-i \lambda_{ab} \hat{a}_1^{\dagger} \hat{a}_1 \hat{b}_1^{\dagger} \hat{b}_1} \nonumber \\
          &=& e^{-i \lambda_{aa} \hat{a}_1^{\dagger} \hat{a}_1} 
              e^{i \lambda_{ab} \hat{a}_1^{\dagger} \hat{a}_1 \hat{b}_1^{\dagger} \hat{b}_1 } \, \hat{a}_1 \,
          e^{-i \lambda_{ab} \hat{a}_1^{\dagger} \hat{a}_1 \hat{b}_1^{\dagger} \hat{b}_1} \nonumber \\
          &=& e^{-i \lambda_{aa} \hat{a}_1^{\dagger} \hat{a}_1} e^{-i \lambda_{ab} \hat{b}_1^{\dagger} \hat{b}_1} \, \hat{a}_1
\label{eqa2}
\end{eqnarray}
Similarly, from permutation symmetry, we have
\begin{equation}
\hat{b}_2 = e^{-i \lambda_{bb} \hat{b}_1^{\dagger} \hat{b}_1} e^{-i \lambda_{ab} \hat{a}_1^{\dagger} \hat{a}_1} \, \hat{b}_1.     
\label{eqb2}
\end{equation}
Finally, to obtain $\hat{a}_3$ and $\hat{b}_3$ we use (\ref{eqa1}) and (\ref{eqb1}) with the phase factor attached to the $\hat{b}$ operator to obtain
\begin{eqnarray}
\hat{a}_3 &=& \cos (\Omega (t_3-t_2)) \hat{a}_2 -i e^{i\phi} \sin (\Omega (t_3-t_2)) \hat{b}_2 \label{eqa3} \\
\hat{b}_3 &=& \cos (\Omega (t_3-t_2)) \hat{b}_2 - i e^{-i\phi} \sin (\Omega (t_3-t_2)) \hat{a}_2 \label{eqb3}
\end{eqnarray} 
We are now in a position to evaluate the number variance of the system throughout the period of the final coupling. The number variances of the fields are defined as
\begin{eqnarray}
{\mathrm{Var}}[N_a] &=& \langle \hat{a}^{\dagger}_3 \hat{a}_3 \hat{a}^{\dagger}_3 \hat{a}_3 \rangle - \langle \hat{a}^{\dagger}_3 \hat{a}_3 \rangle ^2 \label{eqNavariance} \\
{\mathrm{Var}}[N_b] &=& \langle \hat{b}^{\dagger}_3 \hat{b}_3 \hat{b}^{\dagger}_3 \hat{b}_3 \rangle - \langle \hat{b}^{\dagger}_3 \hat{b}_3 \rangle ^2.
\label{eqNbvariance}
\end{eqnarray}
Making use of (\ref{eqa3}) and using the shorthand notation 
\begin{eqnarray}
s &=& \sin[\Omega (t_3 - t_2)] \\
c &=& \cos[\Omega (t_3 - t_2)]
\end{eqnarray}
we have
\begin{eqnarray}
\hat{a}^{\dagger}_3 \hat{a}_3 &=& \hat{a}^{\dagger}_2 \hat{a}_2 c^2 +  \hat{b}^{\dagger}_2 \hat{b}_2 s^2 + i c s (e^{-i \phi} \hat{b}^{\dagger}_2 \hat{a}_2 - e^{i \phi} \hat{a}^{\dagger}_2 \hat{b}_2) \\
\hat{a}^{\dagger}_3 \hat{a}_3 \hat{a}^{\dagger}_3 \hat{a}_3 &=& \hat{a}^{\dagger}_2 \hat{a}_2 \hat{a}^{\dagger}_2 \hat{a}_2 c^4 + \hat{b}^{\dagger}_2 \hat{b}_2 \hat{b}^{\dagger}_2 \hat{b}_2 s^4 \nonumber \\
&& + c^2 s^2 [ \hat{a}^{\dagger}_2 \hat{a}_2 + \hat{b}^{\dagger}_2 \hat{b}_2 + 4 \hat{a}^{\dagger}_2 \hat{a}_2 \hat{b}^{\dagger}_2 \hat{b}_2 -e^{2 i \phi} \hat{a}^{\dagger}_2 \hat{a}^{\dagger}_2 \hat{b}_2 \hat{b}_2 -e^{-2 i \phi} \hat{b}^{\dagger}_2 \hat{b}^{\dagger}_2 \hat{a}_2 \hat{a}_2 ] \nonumber \\
&& + i c^3 s [2 e^{-i \phi} \hat{b}^{\dagger}_2 \hat{a}^{\dagger}_2 \hat{a}_2 \hat{a}_2 - 2 e^{i \phi} \hat{a}^{\dagger}_2 \hat{a}^{\dagger}_2 \hat{a}_2 \hat{b}_2 + e^{-i \phi} \hat{b}^{\dagger}_2 \hat{a}_2 - e^{i \phi} \hat{a}^{\dagger}_2 \hat{b}_2 ] \nonumber \\
&& + i c s^3 [2 e^{-i \phi} \hat{b}^{\dagger}_2 \hat{b}^{\dagger}_2 \hat{a}_2 \hat{b}_2 - 2 e^{i \phi} \hat{a}^{\dagger}_2 \hat{b}^{\dagger}_2 \hat{b}_2 \hat{b}_2 + e^{-i \phi} \hat{b}^{\dagger}_2 \hat{a}_2 - e^{i \phi} \hat{a}^{\dagger}_2 \hat{b}_2 ]
\end{eqnarray}
\begin{eqnarray}
\hat{b}^{\dagger}_3 \hat{b}_3 &=& \hat{b}^{\dagger}_2 \hat{b}_2 c^2 +  \hat{a}^{\dagger}_2 \hat{a}_2 s^2 + i c s ( e^{i \phi} \hat{a}^{\dagger}_2 \hat{b}_2 - e^{-i \phi} \hat{b}^{\dagger}_2 \hat{a}_2) \\
\hat{b}^{\dagger}_3 \hat{b}_3 \hat{b}^{\dagger}_3 \hat{b}_3 &=& \hat{b}^{\dagger}_2 \hat{b}_2 \hat{b}^{\dagger}_2 \hat{b}_2 c^4 + \hat{a}^{\dagger}_2 \hat{a}_2 \hat{a}^{\dagger}_2 \hat{a}_2 s^4 \nonumber \\
&& + c^2 s^2 [ \hat{b}^{\dagger}_2 \hat{b}_2 + \hat{a}^{\dagger}_2 \hat{a}_2 + 4 \hat{b}^{\dagger}_2 \hat{b}_2 \hat{a}^{\dagger}_2 \hat{a}_2 - e^{-2 i \phi} \hat{b}^{\dagger}_2 \hat{b}^{\dagger}_2 \hat{a}_2 \hat{a}_2 -e^{2 i \phi} \hat{a}^{\dagger}_2 \hat{a}^{\dagger}_2 \hat{b}_2 \hat{b}_2 ] \nonumber \\
&& + i c^3 s [2 e^{i \phi} \hat{a}^{\dagger}_2 \hat{b}^{\dagger}_2 \hat{b}_2 \hat{b}_2 - 2 e^{-i \phi} \hat{b}^{\dagger}_2 \hat{b}^{\dagger}_2 \hat{b}_2 \hat{a}_2 + e^{i \phi} \hat{a}^{\dagger}_2 \hat{b}_2 - e^{-i \phi} \hat{b}^{\dagger}_2 \hat{a}_2 ] \nonumber \\
&& + i c s^3 [2 e^{i \phi} \hat{a}^{\dagger}_2 \hat{a}^{\dagger}_2 \hat{b}_2 \hat{a}_2 - 2 e^{-i \phi} \hat{b}^{\dagger}_2 \hat{a}^{\dagger}_2 \hat{a}_2 \hat{a}_2 + e^{i \phi} \hat{a}^{\dagger}_2 \hat{b}_2 - e^{-i \phi} \hat{b}^{\dagger}_2 \hat{a}_2 ] 
\end{eqnarray}
Clearly, to calculate (\ref{eqNavariance}) and (\ref{eqNbvariance}) we will need the expectation values of terms quartic and quadratic in $\hat{a}_2$ and ${\hat{b}_2}$. Specifically, we require $\langle \hat{a}^{\dagger}_2 \hat{a}_2 \rangle$, $\langle \hat{b}^{\dagger}_2 \hat{b}_2 \rangle$,  $\langle \hat{a}^{\dagger}_2 \hat{b}_2 \rangle$, $\langle \hat{a}^{\dagger}_2 \hat{a}_2 \hat{a}^{\dagger}_2 \hat{a}_2 \rangle$, $\langle \hat{b}^{\dagger}_2 \hat{b}_2 \hat{b}^{\dagger}_2 \hat{b}_2 \rangle$ , $\langle \hat{a}^{\dagger}_2 \hat{a}_2 \hat{b}^{\dagger}_2 \hat{b}_2 \rangle$, $\langle \hat{a}^{\dagger}_2 \hat{a}^{\dagger}_2 \hat{b}_2 \hat{b}_2 \rangle$, $\langle \hat{a}^{\dagger}_2 \hat{b}^{\dagger}_2 \hat{a}_2 \hat{a}_2 \rangle$, and $\langle \hat{a}^{\dagger}_2 \hat{b}^{\dagger}_2 \hat{b}_2 \hat{b}_2 \rangle$, as well as their Hermitian conjugates.

As $\hat{a}^{\dagger}_1 \hat{a}_1$ and $\hat{b}^{\dagger}_1 \hat{b}_1$ commute with the Hamiltonian $\hat{H}_2$, they are constants of motion so we have
\begin{equation}
\langle \hat{a}^{\dagger}_2 \hat{a}_2 \rangle = \langle \hat{a}^{\dagger}_1 \hat{a}_1 \rangle = \langle \alpha, \beta | \hat{a}^{\dagger}_1 \hat{a}_1 |\alpha, \beta \rangle = |\alpha|^2 = n_a
\end{equation}
\begin{equation}
\langle \hat{b}^{\dagger}_2 \hat{b}_2 \rangle = \langle \hat{b}^{\dagger}_1 \hat{b}_1 \rangle = \langle \alpha, \beta | \hat{b}^{\dagger}_1 \hat{b}_1 |\alpha, \beta \rangle = |\beta|^2 = n_b
\end{equation}
\begin{equation}
\langle \hat{a}^{\dagger}_2 \hat{a}_2 \hat{b}^{\dagger}_2 \hat{b}_2 \rangle = \langle \hat{a}^{\dagger}_1 \hat{a}_1 \hat{b}^{\dagger}_1 \hat{b}_1 \rangle = \langle \alpha, \beta | \hat{a}^{\dagger}_1 \hat{a}_1 \hat{b}^{\dagger}_1 \hat{b}_1 |\alpha, \beta \rangle = |\alpha|^2 |\beta|^2 = n_a n_b
\end{equation}
Furthermore, $\hat{a}^{\dagger}_2 \hat{a}_2 \hat{a}^{\dagger}_2 \hat{a}_2$ and $\hat{b}^{\dagger}_2 \hat{b}_2 \hat{b}^{\dagger}_2 \hat{b}_2$ also commute with $H_2$, giving
\begin{equation}
\langle \hat{a}^{\dagger}_2 \hat{a}_2 \hat{a}^{\dagger}_2 \hat{a}_2 \rangle = \langle \hat{a}^{\dagger}_1 \hat{a}_1 \hat{a}^{\dagger}_1 \hat{a}_1 \rangle = \langle \alpha, \beta | \hat{a}^{\dagger}_1 \hat{a}^{\dagger}_1 \hat{a}_1 \hat{a}_1 + \hat{a}^{\dagger}_1 \hat{a}_1 |\alpha, \beta \rangle = n_a^2 + n_a
\end{equation}
\begin{equation}
\langle \hat{b}^{\dagger}_2 \hat{b}_2 \hat{b}^{\dagger}_2 \hat{b}_2 \rangle = \langle \hat{b}^{\dagger}_1 \hat{b}_1 \hat{b}^{\dagger}_1 \hat{b}_1 \rangle = \langle \alpha, \beta | \hat{b}^{\dagger}_1 \hat{b}^{\dagger}_1 \hat{b}_1 \hat{b}_1 + \hat{b}^{\dagger}_1 \hat{b}_1 |\alpha, \beta \rangle = n_b^2 + n_b
\end{equation}
The remaining terms are not constants of motion, so we proceed by making use of (\ref{eqa2}) and (\ref{eqb2}). For $\langle \hat{a}^{\dagger}_2 \hat{b}_2 \rangle$ we have
\begin{equation}
\hat{a}^{\dagger}_2 \hat{b}_2 = \hat{a}^{\dagger}_1 e^{-i (\lambda_{bb} - \lambda_{ab}) \hat{b}^{\dagger}_1 \hat{b}_1} e^{i (\lambda_{aa} - \lambda_{ab}) \hat{a}^{\dagger}_1 \hat{a}_1} \hat{b}_1. 
\end{equation}
To calculate the expectation value of this expression, the creation and annihilation operators must be normally ordered. Making use of the identity \cite{louisell}
\begin{equation}
e^{x \hat{a}^{\dagger} \hat{a}} = \sum_{n=0}^{\infty} \frac{(e^x - 1)^n} {n!} (\hat{a}^{\dagger})^n (\hat{a})^n
\end{equation}
allows us to compute expectation values of coherent states of the form
\begin{eqnarray}
\langle \alpha | e^{x \hat{a}^{\dagger} \hat{a}} |\alpha \rangle &=& \sum_{n=0}^{\infty} \frac{(e^x - 1)^n} {n!} |\alpha|^{2n} \nonumber \\
&=& \exp[|\alpha|^2(e^x -1)]. \label{eqExpectationValueOfExponential}
\end{eqnarray}
This gives
\begin{eqnarray}
\langle \hat{a}^{\dagger}_2 \hat{b}_2 \rangle &=& \langle \alpha, \beta | \hat{a}^{\dagger}_1 e^{-i (\lambda_{bb} - \lambda_{ab}) \hat{b}^{\dagger}_1 \hat{b}_1} e^{i (\lambda_{aa} - \lambda_{ab}) \hat{a}^{\dagger}_1 \hat{a}_1} \hat{b}_1 | \alpha, \beta \rangle \nonumber \\
&=& \alpha^* \beta \exp[|\alpha|^2(e^{i (\lambda_{aa} - \lambda_{ab})} -1)] \exp[|\beta|^2 (e^{-i (\lambda_{bb} - \lambda_{ab})} -1)].
\end{eqnarray}

We now consider $\langle \hat{a}^{\dagger}_2 \hat{a}^{\dagger}_2 \hat{b}_2 \hat{b}_2 \rangle$. Using (\ref{eqa2}) and (\ref{eqb2}) and the fact that $\hat{a}$ and $\hat{b}$ commute, we have
\begin{eqnarray}
\hat{a}^{\dagger}_2 \hat{a}^{\dagger}_2 \hat{b}_2 \hat{b}_2 &=& \hat{a}^{\dagger}_1 
  e^{i (\lambda_{aa}-\lambda_{ab}) \hat{a}^{\dagger}_1 \hat{a}_1}
  \hat{b}_1 \hat{a}^{\dagger}_1
  e^{i (\lambda_{ab}-\lambda_{bb}) \hat{b}^{\dagger}_1 \hat{b}_1}
  e^{i (\lambda_{aa}-\lambda_{ab}) \hat{a}^{\dagger}_1 \hat{a}_1}
  \hat{b}_1 \\
&=& \hat{a}^{\dagger}_1 \hat{a}^{\dagger}_1 
    e^{2i (\lambda_{aa}-\lambda_{ab}) \hat{a}^{\dagger}_1 \hat{a}_1}
    e^{2i (\lambda_{ab}-\lambda_{bb}) \hat{b}^{\dagger}_1 \hat{b}_1}
    \hat{b}_1 \hat{b}_1 e^{i(\lambda_{aa}-\lambda_{bb})}
\end{eqnarray}
where we have made multiple uses of the identity 
\begin{equation}
\hat{a} e^{i x \hat{a}^{\dagger} \hat{a}} = e^{i x \hat{a}^{\dagger} \hat{a}} \hat{a} e^{ix}
\end{equation}
which follows from (\ref{eqefeidentity}). Finally, applying (\ref{eqExpectationValueOfExponential}) we obtain
\begin{equation}
\langle \hat{a}^{\dagger}_2 \hat{a}^{\dagger}_2 \hat{b}_2 \hat{b}_2 \rangle = \alpha^{*2} \beta^2 e^{i(\lambda_{aa}-\lambda_{bb})} \exp[|\alpha|^2(e^{2i (\lambda_{aa} - \lambda_{bb})} -1)] \exp[|\beta|^2(e^{-2i (\lambda_{bb} - \lambda_{bb})} -1)].
\end{equation}
Using the same techniques one can show that
\begin{eqnarray}
\langle \hat{a}^{\dagger}_2 \hat{b}^{\dagger}_2 \hat{a}_2 \hat{a}_2 \rangle &=& |\alpha|^2 \alpha \beta^* e^{-i(\lambda_{aa}-\lambda_{ab})} \exp[|\alpha|^2(e^{-i (\lambda_{aa} - \lambda_{ab})} -1)] \exp[|\beta|^2(e^{i (\lambda_{bb} - \lambda_{ab})} -1)] \\
\langle \hat{a}^{\dagger}_2 \hat{b}^{\dagger}_2 \hat{b}_2 \hat{b}_2 \rangle &=& |\beta|^2 \alpha^{*} \beta e^{-i(\lambda_{bb}-\lambda_{ab})} \exp[|\alpha|^2(e^{i (\lambda_{aa} - \lambda_{ab})} -1)] \exp[|\beta|^2(e^{-i (\lambda_{bb} - \lambda_{ab})} -1)].
\end{eqnarray}
Employing the definitions (\ref{eqsDef})--(\ref{eqDDef}) we obtain
\begin{eqnarray}
\langle \hat{a}^{\dagger}(t_3) \hat{a}(t_3) \rangle &=& n_a c^2 + n_b s^2 + i c s (A B^* - A^* B) \\
\langle \hat{b}^{\dagger}(t_3) \hat{b}(t_3) \rangle &=& n_a s^2 + n_b c^2 - i c s (A B^* - A^* B) \\
\langle \hat{a}^{\dagger}(t_3) \hat{a}(t_3) \hat{a}^{\dagger}(t_3)  \hat{a}(t_3) \rangle &=& (n_a^2 + n_a) c^4 + (n_b^2 + n_b) s^4 \nonumber \\
&& + c^2 s^2 [n_a +n_b +4 n_a n_b - e^{i(\lambda_{aa}-\lambda_{bb})} B_2 A_2^* - e^{-i(\lambda_{aa}-\lambda_{bb})} B_2^* A_2] \nonumber \\
&& + i c^3 s [2 n_a e^{-i(\lambda_{aa}-\lambda_{ab})}] A B^* - 2 n_a e^{i(\lambda_{aa}-\lambda_{ab})} A^* B + A B^* - A^* B] \nonumber \\
&& + i c s^3 [2 n_b e^{i(\lambda_{bb}-\lambda_{ab})}] A B^* - 2 n_b e^{-i(\lambda_{bb}-\lambda_{ab})} A^* B + A B^* - A^* B] \\
\langle \hat{b}^{\dagger}(t_3)  \hat{b}(t_3) \hat{b}^{\dagger}(t_3)  \hat{b}(t_3) \rangle &=& (n_b^2 + n_b) c^4 + (n_a^2 + n_a) s^4 \nonumber \\
&& + c^2 s^2 [n_a +n_b +4 n_a n_b - e^{-i(\lambda_{aa}-\lambda_{bb})} B_2^* A_2 - e^{i(\lambda_{aa}-\lambda_{bb})} B_2 A_2^*] \nonumber \\
&& + i c^3 s [2 n_b e^{-i(\lambda_{bb}-\lambda_{ab})}] A^* B - 2 n_b e^{i(\lambda_{bb}-\lambda_{ab})} A B^* - A^* B + A B^*] \nonumber \\
&& + i c s^3 [2 n_a e^{i(\lambda_{aa}-\lambda_{ab})}] A^* B - 2 n_a e^{-i(\lambda_{aa}-\lambda_{ab})} A B^* - A^* B + A B^*]. 
\end{eqnarray}
When substituted into (\ref{eqNavariance}) and (\ref{eqNbvariance}) these expectation values give the absolute number variances stated in Section \ref{secTwoModeAnalytic}. To calculate the number difference variance, we note
\begin{equation}
{\mathrm{Var}}[N_a-N_b] = {\mathrm{Var}}[N_a] + {\mathrm{Var}}[N_b] - 2\langle \hat{a}^{\dagger}_3 \hat{a}_3 \hat{b}^{\dagger}_3 \hat{b}_3\rangle + 2\langle \hat{a}^{\dagger}_3  \hat{a}_3 \rangle \langle \hat{b}^{\dagger}_3 \hat{b}_3\rangle \label{eqNumDiffVarianceAppendix}.
\end{equation}
We write (\ref{eqNumDiffVarianceAppendix}) in terms of $\hat{a}_2$ and $\hat{b}_2$ by using using (\ref{eqa3}) and (\ref{eqb3}), and then use the expectation values calculated earlier in this Appendix. This procedure gives the expression (\ref{eqNumDiffVariance}) given in Section \ref{secTwoModeAnalytic}.

\section{Derivation of Bogoliubov squeezing formula}
\label{appendixBogDerivation}

We consider the Bogoliubov model presented in Section~\ref{sec:2D3Dsqueezing} taken in the limit of a homogeneous condensate with $\chi_{aa}\neq 0$, $\chi_{ab}=\chi_{bb}=0$.  In this limit, the Hamiltonian for the internal state $\ket{a}$ during the hold time is
\begin{align}
  \hat{H}_{a} &= \int \left[ \hat{\psi}_a^\dagger \left(-\frac{\hbar^2}{2 m}\nabla^2\right) \hat{\psi}_a + \frac{U_{aa}}{2} \hat{\psi}_a^\dagger \hat{\psi}_a^\dagger \hat{\psi}_a \hat{\psi}_a \right] \,dV. \label{eqBogOriginalHamiltonian}
\end{align}
To find the response of the condensate to small fluctuations we define fluctuation operators for the non-zero momentum modes $\delta\hat{a}_\mathbf{k} = \hat{a}_\mathbf{k} - \alpha_\mathbf{k}$ where $\alpha_\mathbf{k} = \expect{\hat{a}_\mathbf{k}}$ is the condensate mean field ($\alpha_\mathbf{k}=0$ for $\mathbf{k} \neq 0$ as the condensate is homogenous).  The Hamiltonian \eqref{eqBogOriginalHamiltonian} is then expanded in terms of $\delta\hat{a}_\mathbf{k}$ retaining only the lowest order contributions.  The lowest order contribution is second order:
\begin{align}
  \hat{H}_{a}^{(2)} &= \sum_{\mathbf{k}\neq 0} \left[ \frac{\hbar^2 \mathbf{k}^2}{2M} \delta\hat{a}_\mathbf{k}^\dagger \delta\hat{a}_\mathbf{k} + \frac{1}{2} \hbar \chi_{aa} n_a \left(\delta\hat{a}_\mathbf{k}^\dagger \delta\hat{a}_{-\mathbf{k}}^\dagger + \delta\hat{a}_{-\mathbf{k}} \delta\hat{a}_{\mathbf{k}} \right) \right].
\end{align}
This Hamiltonian is diagonalised with the standard Bogoliubov transformation \cite{PethickSmith}
\begin{align}
  \delta \hat{a}_\mathbf{k}(t) &= u_\mathbf{k} \delta\hat{b}_\mathbf{k}(t) - v_k \delta\hat{b}_{-\mathbf{k}}^\dagger(t),\\
  u_k^2 &= v_k^2+1 = \frac{1}{2} \left(\frac{\omega^0_\mathbf{k} + \chi_{aa}n_a}{\omega_\mathbf{k}} +1\right),
\end{align}
where $\omega^0_\mathbf{k} = \hbar \mathbf{k}^2/2M$ is the frequency of a free particle, and $\omega_\mathbf{k} = \sqrt{\omega^0_\mathbf{k}(\omega^0_\mathbf{k} + 2 \chi_{aa} n_a)}$ are the frequencies of the Bogoliubov modes:
\begin{align}
  \delta\hat{b}_\mathbf{k}(t) &= \exp\left(-i \omega_\mathbf{k} t\right)\delta\hat{b}_\mathbf{k}(0).
\end{align}

The expectation value of the number of atoms in momentum mode $\hbar\mathbf{k}$ after the hold time is then found to be
\begin{align}
  \expect{\hat{n}_\mathbf{k}(t_2)} &= \expect{\delta\hat{a}^\dagger_\mathbf{k}(t_2) \delta\hat{a}_\mathbf{k}(t_2)} \\
  \begin{split}
  &=\left[1 + 4 u_\mathbf{k}^2(2 u_\mathbf{k}^2 - 1)\sin^2(\omega_\mathbf{k} \tau_\text{hold})\right] \expect{\delta \hat{a}^\dagger_\mathbf{k}(t_1) \delta\hat{a}_\mathbf{k}(t_1)} \\
  &\phantom{\mathrel{=}}- 4 u_\mathbf{k} v_\mathbf{k} \sin(\omega_\mathbf{k}\tau_\text{hold}) \Im\left\{\left[1-i2 u_\mathbf{k}^2 \sin(\omega_k \tau_\text{hold})\right] \expect{\delta\hat{a}_{-\mathbf{k}}(t_1) \delta\hat{a}_\mathbf{k}(t_1)}\right\}\\
  &\phantom{\mathrel{=}} + 4 u_\mathbf{k}^2 v_\mathbf{k}^2 \sin^2(\omega_\mathbf{k}\tau_\text{hold}).
  \end{split}
\end{align}
If the non-zero momentum modes are in a vacuum state at $t=t_1$ then this reduces to
\begin{align}
  \expect{\hat{n}_\mathbf{k}(t_2)} &= 4 u_\mathbf{k}^2 v_\mathbf{k}^2 \sin^2(\omega_\mathbf{k}\tau_\text{hold}) \\
  &= \left[n_a \chi_{aa} \tau_\text{hold} \sinc(\omega_\mathbf{k} \tau_\text{hold}) \right]^2.
\end{align}
After the final Rabi coupling pulse the occupation of the non-zero momentum mode $\hbar\mathbf{k}$ is
\begin{align}
  n_a(\mathbf{k}, t_3) &= \left[n_a \chi_{aa} \tau_\text{hold} \cos(\theta)\sinc(\omega_\mathbf{k} \tau_\text{hold}) \right]^2
\end{align}
which is the expression given earlier in Eq.~\eqref{eqBogOccupation}.

\end{document}